%% file: main.tex
\documentclass{article}
\pdfoutput=1
\PassOptionsToPackage{numbers, compress}{natbib}
\usepackage[preprint]{neurips_2023}
\usepackage{tcolorbox}
\definecolor{cadmiumgreen}{rgb}{0.0, 0.42, 0.24}
\definecolor{oldmauve}{rgb}{0.4, 0.19, 0.28}
\definecolor{royalazure}{rgb}{0.0, 0.22, 0.66}
\definecolor{harvardcrimson}{rgb}{0.79, 0.0, 0.09}
\definecolor{lightmauve}{rgb}{0.86, 0.82, 1.0}
\definecolor{darkbrown}{rgb}{0.4, 0.26, 0.13}
\usepackage{hyperref}
\hypersetup{colorlinks}
\usepackage[utf8]{inputenc} %
\usepackage[T1]{fontenc}    %
\usepackage{url}            %
\usepackage{booktabs}       %
\usepackage{amsfonts}       %
\usepackage{nicefrac}       %
\usepackage{microtype}      %
\usepackage{xcolor}         %
\usepackage{bm}
\usepackage{graphicx}
\usepackage{amsmath}
\usepackage{amssymb}
\usepackage{booktabs}
\usepackage{algorithm}
\usepackage{algpseudocode}
\usepackage{diagbox}
\usepackage{float}
\usepackage{multirow}
\usepackage{subcaption}
\usepackage{pifont}
\usepackage{wrapfig}
\usepackage[font=small]{caption}

\newcommand{\minus}{\scalebox{0.75}[1.0]{$-$}}

\long\def\comment#1{}

\usepackage[capitalize]{cleveref}
\crefname{section}{Sec.}{Secs.}
\Crefname{section}{Section}{Sections}
\Crefname{table}{Table}{Tables}
\crefname{table}{Tab.}{Tabs.}
\crefname{algorithm}{Algorithm}{Algorithm}

\newcommand*{\affaddr}[1]{#1} 
\newcommand*{\affmark}[1][*]{\textsuperscript{#1}}
\newcommand*{\email}[1]{\texttt{#1}}

\def\ie{$i.e.$} 
\def\eg{$e.g.$} 
\def\etc{$etc.$} 
\def\wrt{$w.r.t.~$}

\def\gD{{\mathcal{D}}}

\def\gL{{\mathcal{L}}}

\def\gX{{\mathcal{X}}}
\def\gY{{\mathcal{Y}}}

\def\vm{{\bm{m}}}

\def\vx{{\bm{x}}}

\def\mH{{\bm{H}}}

\begin{document}
\title{Boosting Backdoor Attack with A Learnable Poisoning Sample Selection Strategy}
\author{%
Zihao Zhu\affmark[1]\quad
Mingda Zhang\affmark[1] \quad
Shaokui Wei\affmark[1] \\
\textbf{Li Shen}\affmark[2] \quad
\textbf{Yanbo Fan}\affmark[3] \quad
\textbf{Baoyuan Wu}\affmark[1]\thanks{Corresponding Author} 
\\
\affaddr{\affmark[1]School of Data Science, Shenzhen Research Institute of Big Data, \\ The Chinese University of Hong Kong, Shenzhen}\\
\affaddr{\affmark[2]JD Explore Academy}\quad
\affaddr{\affmark[3]Tencent AI Lab}
\\
\small\email{\{zihaozhu, mingdazhang, shaokuiwei\}@link.cuhk.edu.cn; }\\
\small\email{ \{mathshenl, fanyanbo0124\}@gmail.com; wubaoyuan@cuhk.edu.cn}
}
\maketitle
\input{0_abstract}

\input{1_introduction}

\input{2_related}

\input{3_method}

\input{4_experiment}

\input{5_analysis}

\input{6_conclusion}
\clearpage
\bibliographystyle{plainnat}
\bibliography{main}
\end{document}

%% file: 0_abstract.tex
\begin{abstract} 
Data-poisoning based backdoor attacks aim to insert backdoor into models by manipulating training datasets without controlling the training process of the target model.
Existing attack methods mainly focus on designing triggers or fusion strategies between triggers and benign samples. However, they often randomly select samples to be poisoned, disregarding the varying importance of each poisoning sample in terms of backdoor injection.
A recent selection strategy filters a fixed-size poisoning sample pool by recording forgetting events, but it fails to consider the remaining samples outside the pool from a global perspective. Moreover, computing forgetting events requires significant additional computing resources. 
Therefore, how to efficiently and effectively select poisoning samples from the entire dataset is an urgent problem in backdoor attacks.
To address it, firstly, we introduce a poisoning mask into the regular backdoor training loss. 
We suppose that a backdoored model training with hard poisoning samples has a more backdoor effect on easy ones, which can be implemented by hindering the normal training process (\ie, maximizing loss \wrt mask). 
To further integrate it with normal training process, we then propose a learnable poisoning sample selection strategy to learn the mask together with the model parameters through a min-max optimization.
Specifically, the outer loop aims to achieve the backdoor attack goal by minimizing the loss based on the selected samples, while the inner loop selects hard poisoning samples that impede this goal by maximizing the loss.
After several rounds of adversarial training, we finally select effective poisoning samples with high contribution.
Extensive experiments on benchmark datasets demonstrate the effectiveness and efficiency of our approach in boosting backdoor attack performance.
\end{abstract}

%% file: 1_introduction.tex
\section{Introduction}
\label{sec:intro}

Training large-scale deep neural networks (DNNs) often requires massive training data. Considering the high cost of collecting or labeling massive training data, users may resort to downloading publicly free data from an open-sourced repository or buying data from a  third-party data supplier. 
However, these unverified data may expose the model training to a serious threat of data-poisoning based backdoor attacks. Through manipulating a few training samples, the adversary could insert the malicious backdoor into the trained model, which performs well on benign samples, but will predict any poisoned sample with trigger as the target class.

Several seminal backdoor attack methods (\eg~BadNets~\cite{badnet}, Blended~\cite{blended}, SSBA~\cite{ssba}, SIG~\cite{sig}, TrojanNN~\cite{trojannn} \etc) have shown good attack performance (\ie, high attack success rate while keeping high clean accuracy) on mainstream DNNs. 
Most of these attack methods focus on designing diverse triggers (\eg~patch trigger~\cite{badnet}, or signal trigger~\cite{sig}), or the fusion strategy of inserting the trigger into the benign sample (\eg~alpha-blending adopted in Blended~\cite{blended}, digital steganography adopted in SSBA~\cite{ssba}), to make the poisoned samples stealthy and effective. 
However, it is often assumed that a few benign samples are randomly selected from the benign training dataset to generate poisoned samples. 
Some recent works~\cite{influence,katharopoulos2018not,paul2021deep} suggest that not all data are equally useful for training DNNs --- some have greater importance for the task at hand or are more rich in informative content than others. 
Several selection strategies, such as uncertainty-based~\cite{coleman2019selection}, influence function~\cite{influence}, forgetting events~\cite{forgetting}, have been proposed to mine important samples for coreset selection~\cite{borsos2020coresets,killamsetty2021glister,killamsetty2021grad,mirzasoleiman2020coresets,killamsetty2021retrieve}, data valuation~\cite{yoon2020data,nohyundata,just2023lava} and active learning~\cite{sener2017active,kaushal2019learning,chang2017active}.

It inspires us to explore whether the backdoor performance could be boosted if the samples to be poisoned are selected according to some strategies rather than randomly, especially depending on the trigger and benign data. 
This underappreciated problem has rarely been studied in the backdoor learning community, and there is only one attempt~\cite{ijcai2022forget} try to solve it. 
A filtering-and-updating strategy (FUS)~\cite{ijcai2022forget} has been proposed to filter poisoning samples within a fixed-size sample pool based on forgetting events~\cite{forgetting}, while disregarding the remaining samples beyond the pool, which is a local perspective.
Besides, computing forgetting events for each updating step requires the same number of epochs as the full training process, resulting in a significant increase in computational cost, which is impractical in real-world scenarios.
Hence, {how to efficiently and effectively select samples to be poisoned with a global perspective from the entire dataset, while maintaining general applicability to diverse backdoor attacks} is still an urgent problem to be solved.

To address the aforementioned issue,  we propose a \textbf{L}earnable \textbf{P}oisoning sample \textbf{S}election strategy (LPS) that depends on triggers, poisoned fusion strategies, and benign data. 
The key idea behind it is that if we can successfully implant the backdoor into the model through hard poisoning samples, the backdoor behavior can be effectively generalized to other easier samples at the inference stage.
A learnable binary poisoning mask $\vm$ is first introduced into the regular backdoor training loss (\cref{eq:loss2}).
Then finding hard samples can intuitively be obtained by hindering backdoor training process (\ie, maximize loss \wrt $\vm$).
In order to further fuse it with normal backdoor training, we consequently formulate the poisoning sample selection as a min-max optimization via an adversarial process.
During the min-max two-player game, the inner loop optimizes the  mask to identify hard poisoning sample, while the outer loop optimizes the  model's parameters to train a backdoored model based on the selected  samples.
By adversarially training the min-max problem over multiple rounds, we finally obtain the high-contributed poisoning samples  that serve  the malicious backdoor objective. 
The proposed LPS strategy can be naturally adopted in any off-the-shelf data-poisoning based backdoor attacks. 
Extensive evaluations with state-of-the-art backdoor attacks are conducted on benchmark datasets. The results demonstrate the superiority of our LPS strategy over both the random selection and the FUS strategy \cite{ijcai2022forget}, while resulting in significant computational savings.

The main contributions of this work are three-fold. 
\textbf{1)} We propose a general  backdoor training loss that incorporates a binary poisoning mask.
\textbf{2)} We propose a learnable poisoning sample selection strategy by formulating it as a min-max optimization problem. 
\textbf{3)} We provide extensive experiments to verify the effectiveness of the proposed selection strategy on significantly boosting existing data-poisoning backdoor attacks.

%% file: 2_related.tex
\section{Related work}
\vspace{-1mm}
\textbf{Backdoor attack.}
According to the threat model, existing backdoor attacks can be partitioned into two categories: \textit{data-poisoning based}~\cite{badnet,blended,ssba,wanet,sig,trojannn,wang2023robust} and \textit{training-controllable based}~\cite{nguyen2020input,doan2021backdoor,doan2021lira,wang2022bppattack}. 
In this work, we focus on the former threat model, where the adversary can only manipulate the training dataset and the training process is inaccessible. 
Thus, here we mainly review the related data-poisoning based attacks, and we refer readers to recent surveys \cite{wu2023adversarial,li2020backdoor,wu2022backdoorbench} for a detailed introduction to training-controllable attacks. 
BadNets~\cite{badnet} was the first attempt to stamp a patch on the benign image as the poisoned image, revealing the existence of backdoor in deep learning. 
Blended ~\cite{blended} used the alpha blending strategy to make the trigger invisible to evade human inspection.
SIG~\cite{sig} generated a ramp or triangle signal  as the trigger. 
TrojanNN attack~\cite{trojannn} optimized the trigger by maximizing its activation on selected neurons related.
SSBA~\cite{ssba} adopted a digital stenography  to fuse a specific string into images by autoencoder, to generate sample-specific triggers.
Subsequently, more stealthy and effective attacks~\cite{zeng2021rethinking,zhang2022poison,salem2022dynamic,turner2019label,souri2021sleeper,nguyen2020input,doan2022marksman} have been successively proposed.
Meanwhile, some defense methods~\cite{tran2018spectral,huang2022backdoor,wang2023unicorn,chai2022oneshot,ac,doan2023bdvits,zeng2023sift,zeng2022narcissus,zhu2023enhancing} have been proposed as shields to resist attacks.
The commonality of  above attacks is that they focused on designing triggers or the fusion strategy, while overlooking how to select benign samples to generate poisoned samples, and simply adopted the random selection strategy.  
Instead, we aim to boost existing data-poisoning backdoor attacks through a learnable poisoning sample selection strategy depending on the trigger and benign data. 
The filtering step is based on the forgetting events \cite{forgetting} recorded on a small number of adversaries, which ensures that the differences between samples can emerge.
Afterwards, some new poisoned samples are sampled randomly from the candidate set to update the pool. The above two steps are iterated several times to find a suitable solution.

\textbf{Poisoning sample selection in backdoor attack.}
To the best of our knowledge,  there is only one work~\cite{ijcai2022forget} focusing on poisoning sample selection for backdoor attack. 
A filtering-and-updating strategy (FUS) has been proposed in~\cite{ijcai2022forget}  to iteratively filter and update a sample pool. The filtering step filters easily forgotten poisoning samples based forgetting events~\cite{forgetting}, which are recorded by the same number of epochs as the full training process.
Afterwards, some new poisoned samples are sampled randomly from the candidate set to update the pool.
The above two steps are iterated several times to find a suitable solution.
As the pioneering work, FUS shows good improvement in backdoor effect compared to the random selection strategy. However, FUS requires tens of times more computing resources,  which is not acceptable in practice.

%% file: 3_method.tex
\begin{figure*}[!t] 
\centering 
\includegraphics[width=0.95\linewidth]{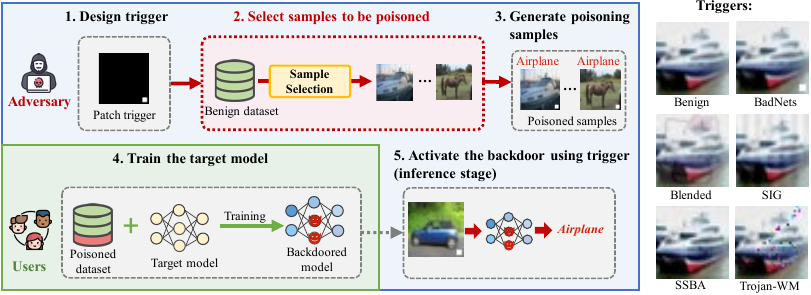}
\vspace{-2mm}
\caption{The general procedure of data-poisoning based backdoor attack and examples of representative triggers.} 
\label{fig:pipeline} 
\vspace{-5mm}
\end{figure*}

\section{Preliminary}

\textbf{Threat model.}
\label{subsec: threat model}
We consider the threat model that the adversary can only manipulate the training dataset with the training process inaccessible, dubbed \textit{data-poisoning based backdoor attack}. It applies to the scenario in which the user trains a neural network based on an unverified dataset.

\textbf{General procedure of data-poisoning based backdoor attacks.}
\label{subsec: general procedure of backdoor attack}
Here we describe the general procedure of data-poisoning based backdoor attacks. 
As shown in \cref{fig:pipeline}, it consists of  5 steps:

\noindent \ding{182} \textbf{Design trigger (by adversary).} The first step of backdoor attack is to design a trigger $\bm\epsilon$, of which the format could be diverse in different applications, such as one image with particular textures in computer vision tasks, as shown in the right part of \cref{fig:pipeline}.

\noindent \ding{183} \textbf{Select samples to be poisoned (by adversary).} 
Let $\gD=\{(\vx_i,y_i)\}_{i=1}^{|\gD|}$ denote the original benign training dataset that contains $|\gD|$ \textit{i.i.d.} samples, where $\vx_i \in \gX$ denotes the input feature, $y_i \in \gY=\{1,\dots,K\}$ is the ground-truth label of $\vx_i$. There are $K$ candidate classes, and the size of class $k$ is denoted as $n_k$. 
For clarity, we assume that all training samples are ranked following the class indices, \ie, $(\text{samples of class $1$}), (\text{samples of class $2$}), \ldots, (\text{samples of}$  $\text{class $K$})$. %
To ensure stealthiness and avoid harm to clean accuracy, the adversary often selects a small fraction of benign samples to be poisoned. 
Here we define a binary vector $\vm=\left[m_1,m_2,\dots,m_{|\gD|}\right]\in \{0,1\}^{|\gD|}$ to represent the poisoning mask, where $m_i=1$ indicates that $\vx_i$ is selected to be poisoned and $m_i=0$ means not selected. 
We denote $\alpha :={\sum_{i=1}^{|\gD|}m_i}{\big /}{|\gD|}$ as the poisoning ratio. Note that most existing backdoor attack methods randomly select $\alpha \cdot |\gD|$ samples to be poisoned.

\noindent \ding{184} \textbf{Generate poisoned samples (by adversary).} 
Given the trigger $\bm\epsilon$ and the selected sample $\vx_i$ (\ie, $m_i = 1$), the adversary will design some strategies to fuse $\bm\epsilon$ into $\vx_i$ to generate the poisoned sample $\tilde{\vx}_i$, \ie, 
$\tilde{\vx}_i = g(\vx_i, \bm\epsilon)$, 
with $g(\cdot, \cdot)$ denoting the fusion operator (\eg~ the alpha-blending used in Blended \cite{blended}).
Besides, the adversary has authority to change the original ground-truth label $y_i$ to the target label $\tilde{y_i}$.
If target labels remain the same for all poisoning samples (\ie, $ \tilde{y_i}=y_t$), it is called \textit{all-to-one} attack. 
If target labels have differnt types (\eg, $ \tilde{y_i}=y_i+1$), it is called \textit{all-to-all} attack.
If adversary does not change the ground-truth label (\ie, $\tilde{y_i}=y_i$), it is called \textit{clean label} attack.
Thus, the generated poisoned training dataset could be denoted as $\tilde{\gD}=\{(\vx_i, y_i)|_{\text{if}~ m_i=0}, ~ \text{or} ~ (\tilde{\vx}_i, \tilde{y_i})|_{\text{if}~ m_i=1} \}_{i=1}^{|\gD|}$.

\noindent \ding{185} \textbf{Train the target model (by user).} 
Given the poisoned training dataset $\tilde{\gD}$, the user trains the target model $f_{\bm{\theta}_t}$ by minimizing the following loss function:
\begin{flalign}
    & \gL(\bm{\theta}_t;\tilde{\gD}) = \frac{1}{|\tilde{\gD}|}\sum_{(\vx, y) \in \tilde{\gD}} \ell(f_{\bm{\theta}_t}(\vx),y)) 
    \label{eq:loss1}
    \\
    \equiv & \gL(\bm{\theta}_t;\gD,\vm,\bm\epsilon,  g)=\frac{1}{|\gD|}\sum_{i=1}^{|\gD|}\Big[(1-m_i)\cdot \ell(f_{\bm{\theta}_t}(\vx_i),y_i)) + m_i\cdot \ell(f_{\bm{\theta}_t}(\tilde{\vx}_i), y_t)\Big],
    \label{eq:loss2}
\end{flalign}
where $\ell(\cdot,\cdot)$ is the loss function for an individual sample, such as cross-entropy loss.
In \cref{eq:loss2}, we extend \cref{eq:loss1} by introducing binary poisoning mask $\bm{m}$ that described in step 2.

\noindent \ding{186} \textbf{Activate the backdoor using the trigger during the inference stage (by the adversary)} Given the trained model $f_{\bm{\theta}_t}$, the adversary expects to activate the injected backdoor using the trigger $\bm\epsilon$, \ie, fooling $f_{\bm{\theta}_t}$ to predict any poisoned sample $g(\vx_i, \bm\epsilon)$ as the target label $\tilde{y_i}$.

\comment{at the end of the section, we should summarize the challenges in this process and how can we solve these challenges via the proposed selection strategy.}

\begin{figure}[!t]
    \centering
    \includegraphics[width=0.8\linewidth]{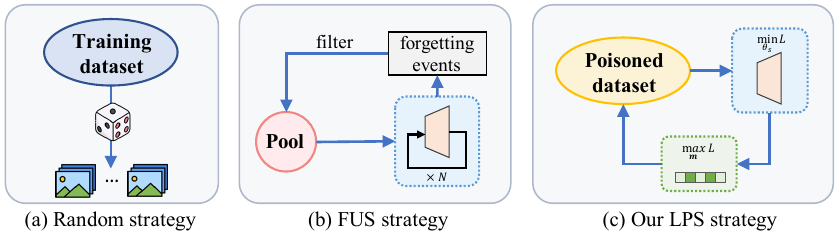}
    \vspace{-2mm}
    \caption{Different poisoning sample selection strategies.}
    \vspace{-5mm}
    \label{fig:method}
\end{figure}

Most backdoor attack methods  concentrate on designing diverse triggers (\ie, step 1) or the fusion strategy (\ie, step 3). 
These attacks typically  randomly select samples for poisoning (\ie, step 2), neglecting the unequal influence of each poisoning samples to the backdoor injection.
Recent FUS strategy~\cite{ijcai2022forget} , as shown in \cref{fig:method}, filters unimportant poisoning samples in a pool based on forgetting events~\cite{forgetting}, while ignoring the rest of the samples outside the pool, which is a local perspective.
Besides, since the characteristics of poisoning samples vary from different attacks, the selected samples that succeed in one attack may not be effective in others.
Therefore, it is a challenging task to develop a poisoning sample selection strategy that can select poisoning samples from the entire dataset and be generally applicable to various backdoor attacks.

\section{Methodology: learnable poisoning sample selection strategy}

This work aims to design a novel sample selection strategy to enhance the impact of a backdoor in the trained target model, denoted as $f_{\boldsymbol{\theta}t}$. 
As the target model $f{\boldsymbol{\theta}t}$ is agnostic to adversaries, we adopt a surrogate model $f{\boldsymbol{\theta}_s}$ as an alternative. 
In order to select poisoning samples from the entire dataset with a global perspective, we opt to directly generate the poisoning mask $\bm{m}$ in step 2. 
We suppose that if backdoor can been implanted into the model through training with \textit{hard} poisoning samples, the backdoor can be generally activated by other \textit{easy} samples during the inference stage.
To achieve this, an intuitive way  is to hinder the normal backdoor training from an opposite direction, \ie, maximize the loss in \cref{eq:loss2} given the surrogate model.
To combine it with the normal training process (\ie, minimize \cref{eq:loss2}), we propose a \textbf{L}earnable \textbf{P}oisoning sample \textbf{S}election (LPS)  strategy to learn the poisoning mask $\vm$ along with the surrogate model's parameters $\boldsymbol{\theta}_s$ through  a min-max optimization:
\begin{equation}
\label{eq:minmax}
    \min_{\bm\theta_s} \max_{\vm \in\{0,1\}^{|\gD|}}  \Big\{\gL(\bm\theta_s, \vm; \gD,\bm\epsilon, g) \quad  \mathrm{s.t.} ~ \mH\vm= \tilde{\alpha}  \cdot \bm\mu\Big\}, 
\end{equation}
where $\gL$ is extended loss including poisoning mask that defined in \cref{eq:loss2}.
$\mH \in \{0,1\}^{K \times |\gD|}$ is defined as: in the $k$-th row,  the entries $\mH(k, \sum_{j=1}^{k-1} n_j + 1: \sum_{j=1}^{k} n_j) = 1$, while other entries are 0. 
$\tilde{\alpha} = \frac{\alpha \cdot |\gD|}{\sum_{k \neq y_t} n_k}$ and $\tilde{\alpha}n_k$ is integer for all $k$.  
$\bm\mu = [\mu_1; \mu_2; \ldots; \mu_K] \in \mathbb{N}^{K}$ is defined as: if $k\neq y_t$, then $\mu_k = n_k$, otherwise $\mu_k = 0$. 
This equation captures three constraints, including: \textbf{1)} $\alpha \cdot |\gD|$ samples are selected to be poisoned; \textbf{2)} the target class samples cannot be selected to be poisoned; \textbf{3)} each non-target class has the same selected ratio $\tilde{\alpha}$ to encourage the diversity of selected samples. %
Note that here we only consider the setting of \textit{all-to-one} attack, but the constraint can be flexibly adjusted for \textit{all-to-all} and \textit{clean label} settings.

\begin{wrapfigure}[14]{r}{0.53\linewidth}
  \centering
  \vspace{-8mm}
  \begin{minipage}{\linewidth}
   \begin{algorithm}[H]
    \renewcommand{\algorithmicrequire}{\textbf{Input:}}
	\renewcommand{\algorithmicensure}{\textbf{Output:}}
	\algnewcommand\algorithmicforeach{\textbf{for each}}
    \algdef{S}[FOR]{ForEach}[1]{\algorithmicforeach\ #1\ \algorithmicdo}
    \caption{ LPS strategy via min-max optimization}
    \label{alg:algorithm}
    \begin{algorithmic}[1]
        \Require  Benign training dataset $\gD$, architecture of the surrogate model $f_{\bm\theta_s}$, maximal iterations $T$, poisoning ratio $\alpha$, trigger $\bm\epsilon$, fusion operator $g$
		\Ensure poisoning mask $\vm$ 
        \State Randomly initialize $\vm_s^{(0)}$, $\bm\theta_s^{(0)}$
        \ForEach{ iteration $t=0$ to $T-1$}
            \State $\triangleright$  Given $\vm^{(t)}$, update $\bm\theta^{(t+1)}_s$   by solving outer 
            \Statex $\quad\quad\ $sub-problem in \cref{eq:outer min}.
            \State $\triangleright$  Given $\bm\theta^{(t+1)}_s$, update $\vm^{(t+1)}$ by solving \Statex $\quad\quad$ inner sub-problem in \cref{eq:inner max1}.
        \EndFor
        \State \Return $\vm_{T}$
    \end{algorithmic}
\end{algorithm}
\end{minipage}
\end{wrapfigure}
\textbf{Remark.}
This min-max objective function (\ref{eq:minmax}) is designed for finding hard poisoning samples with high-contribution for backdoor injection via an adversarial process. 
Specifically, the inner loop encourages to select hard samples for the given model's parameters $\bm\theta_s$ by maximizing the loss \wrt $\vm$, while  the  outer loop aims to update $\bm\theta_s$ by minimizing the loss \wrt $f_{\bm\theta_s}$ to ensure  that a good backdoored model  can be still learned, even based on the hard poisoning mask $\vm$. 
Thus, the two-palyer game between $\vm$ and $\bm\theta_s$ is expected to encourage the selected samples to bring in good backdoor effect, while avoiding over-fitting to the surrogate model $f_{\bm\theta_s}$.

\textbf{Optimization.} 
As summarized in \cref{alg:algorithm}, 
the min-max optimization (\ref{eq:minmax}) could be efficiently solved by alternatively updating $\vm$ and $\bm\theta_s$ as follows:

\ding{169} \textbf{Outer minimization}: given $\vm$, $\bm\theta_s$ could be updated by solving the following sub-problem:
    \begin{flalign}
    \hspace{-1.2em}
      {\bm\theta_s} \in \arg\min_{\bm\theta_s}  ~ \gL(\bm\theta_s; \vm, \gD,\bm\epsilon, g).  
        \label{eq:outer min}
    \end{flalign}
It could be optimized by the standard back-propagation method with stochastic gradient descent (SGD) \cite{SGD}. 
Here we update $\bm\theta_s$ for one epoch in each iteration.

\ding{169} \textbf{Inner maximization}: given $\bm\theta_s$, $\vm$ could be achieved by solving the maximization problem as:
    \begin{flalign}
    \hspace{-1.2em}
    \vm \in    \arg\max_{\vm \in\{0,1\}^{|\gD|}}  \Big\{\gL(\vm; \bm\theta_s, \gD,\bm\epsilon, g), 
       ~ \mathrm{s.t.} ~ \mH\vm= \tilde{\alpha}  \cdot \bm\mu\Big\}.
       \label{eq:inner max1}
    \end{flalign}
    Although it is a constrained binary optimization problem, it is easy to obtain the optimal solution. Specifically, given the hard constraint $\mH\vm= \tilde{\alpha}  \cdot \bm\mu$, the above problem could be separated into $K$ independent sub-problems, \ie, 
    \begin{flalign}
    \hspace{-1em}
        \max_{\vm_k \in\{0,1\}^{n_k}}   
        \frac{1}{|\gD|} \left\{\sum_{i=1}^{|\gD|} \mathbb{I}(y_i=k)\cdot m_i \cdot \big[ \ell(f_{\bm{\theta}_s}(\tilde{\vx}_i), y_t) \!-\!  \ell(f_{\bm{\theta}_s}(\vx_i),y_i) \big], ~ \mathrm{s.t.} ~\mathbf{1}_{n_k}^\top \vm_k \!=\! \tilde{\alpha} \cdot n_k\right\}, 
        \label{eq:inner max2}
    \end{flalign}
    for $\forall k \in \{1,2,\ldots,K\}$ except $k=y_t$. $\vm_k$ denotes the sub-mask vector of $\vm$ corresponding to samples of class $k$, and $\mathbb{I}(a) = 1$ if $a$ is true, otherwise 0. 
    Note that some constant terms \wrt $\vm_k$ have been abandoned in the above sub-problem. And, since it is constrained that only non-target class samples can be selected, $\vm_{y_t}$ is always a zero vector. 
    It is easy to obtain the optimal solution by firstly calculating $\ell(f_{\bm{\theta}_s}(\tilde{\vx}_i), y_t) - \ell(f_{\bm{\theta}_s}(\vx_i),y_i)$ for all samples satisfying $\mathbb{I}(y_i=k)=1$ and ranking them in descending order, then 
    picking the top-($\tilde{\alpha} \cdot n_k)$ indices to set the corresponding $m_i$ as 1, while others as 0.

%% file: 4_experiment.tex
\section{Experiments}

\subsection{Experimental settings}

\textbf{Implementation details.} 
For the training of both surrogate models  and target models, we adopt SGD optimizer with weight decay $5e\minus 4$, the batch size 128, the initial learning rate 0.01 and reduced by 10 after 35 and 55 epochs, respectively. 
The training epoch for target models is 100. 
The maximal iteration $T$  is set as $15$. All experiments are conducted on NVIDIA GTX 3090 GPUs.

\textbf{Datasets and models.} We evaluate on three commonly used  benchmark datasets: CIFAR-10~\cite{cifar}, CIFAR-100~\cite{cifar} and Tiny-ImageNet~\cite{tiny}. The surrogate model and target model are ResNet-18\cite{resnet} and ResNet-34, respectively.

\textbf{Baselines of poisoning sample selection.}
We compare our proposed LPS strategy with two existing poisoning sample selection strategies: \textit{Random} and \textit{FUS}~\cite{ijcai2022forget}.
Random strategy selects benign samples following a uniform distribution. 
FUS~\cite{ijcai2022forget} selects samples according to the sample importance measured by forgetting events\footnote{Note that in the experiments reported in \cite{ijcai2022forget}, FUS appended the generated poisoned samples onto the original benign dataset, rather than replacing the selected benign samples, leading to $|\tilde{\gD}|\ge|\gD|$. To ensure fair comparison, we change it to the traditional setting in existing attacks that the selected benign samples to be poisoned are replaced by the generated samples, thus $|\tilde{\gD}|=|\gD|$.}.
Following the original setting in~\cite{ijcai2022forget}, we set $10$ overall iterations and $60$ epochs for updating the surrogate model in each iteration.

\textbf{Backdoor attacks.}
We consider 5 representative backdoor attacks: 1) visible triggers: BadNets~\cite{badnet}, Blended~\cite{blended}; SIG~\cite{sig}; 2) optimized triggers: Trojan-Watermark (Trojan-WM)~\cite{trojannn}; 3) sample-specific triggers:  SSBA~\cite{ssba}. In addition, we consider 3 poisoning label types: all-to-one, all-to-all and clean label. We visualize different triggers with the same benign image in \cref{fig:pipeline}. The detailed settings of each attack can been found in \textbf{supplement materials}.

\textbf{Backdoor defenses.}
We select 6 representative backdoor defenses to evaluate the resistance of above attack methods with different poisoning sample selection strategies, including Fine-Tuning (FT), Fine-Pruning (FP)~\cite{FP}, Anti-Backdoor Learning (ABL)~\cite{abl}, Channel Lipschitzness Pruning (CLP)~\cite{clp}, Neural Attention Distillation (NAD)~\cite{nad}, Implicit Backdoor Adversarial Unlearning (I-BAU)~\cite{ibau}. The detailed settings of each defense can been found in \textbf{supplement materials}.

\begin{figure}[!t] 
\centering 
\setlength{\abovecaptionskip}{2pt}
\includegraphics[width=0.8\columnwidth]{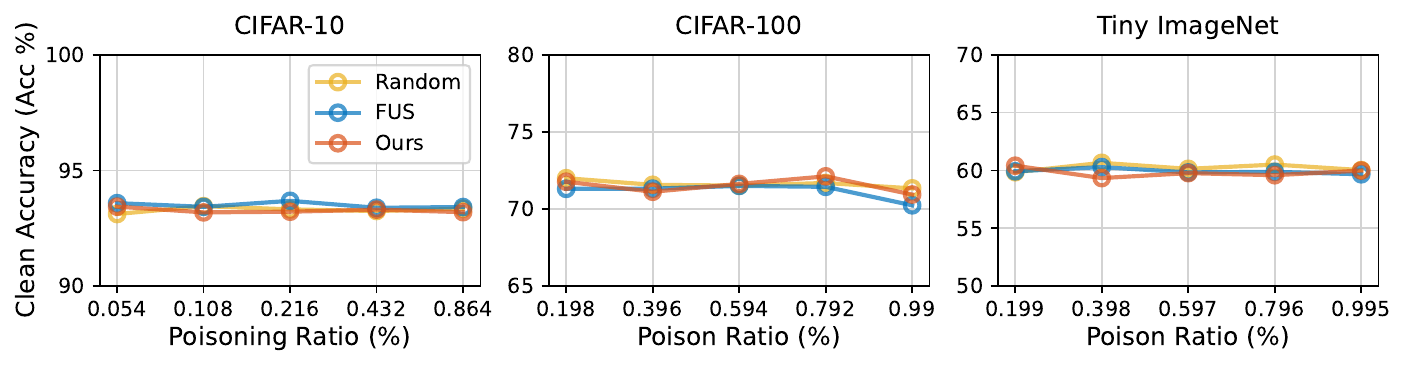}
\caption{Clean accuracy of Blended attack with different backdoor sample selection strategies.} 
\vspace{-6mm}
\label{fig:clean_acc} 
\end{figure}
\input{tables/cifar10_rn18_rn34}

\subsection{Main results}
We evaluate  our  LPS strategy under various experimental settings, including comparisons with baseline strategies on various attacks and poisoning ratios, comparisons on different datasets and resistance to defenses. 
The attack results on CIFAR-10, CIFAR-100, and Tiny-ImageNet can be found in Tab. \ref{tab:cifar10_rn18_rn34},\ref{tab:cifar100_rn18_rn34},\ref{tab:tiny_rn18_rn34} respectively. Additionally, \cref{tab:defense}  presents the defense results on CIFAR-10. 
Besides, we find that due to the low poisoning ratios, the impacts of different poisoning sample selection strategies on the clean accuracy are almost similar (as shown in \cref{fig:clean_acc}).
Thus, for clarity, we omit ACC in most result tables, except for \cref{tab:defense}. Three random trials are conducted for the main experiments  to report  the mean and standard deviation.
More results about different models can be found in  \textbf{supplement materials}.

\input{tables/cifar100_rn18_rn34}
\input{tables/tiny_rn18_rn34}

\textbf{Compare with state-of-the-art baselines.} 
To verify the effectiveness of our proposed LPS strategy, we first compare with two existing strategies on CIFAR-10, in which the surrogate model is ResNet-18 and the target model is ResNet-34.
Different from~\cite{ijcai2022forget}, we conduct experiments under low poisoning ratios ($< 1\%$), which is more stealthy and more likely to escape human inspection.
The attack success rate is shown in \cref{tab:cifar10_rn18_rn34}, where \textit{\#Img/Cls} denotes the number of samples to be poisoned per class for all-to-one setting, and  \textit{pratio} is short for poisoning ratio.
\textbf{1) From a global view}, we observe that LPS strategy outperforms the baselines under most of the settings. 
For example, with $0.216\%$ poisoning ratio, LPS strategy can boost BadNets (all-to-all) by $30.61\%$ compared to FUS, and Blended (all-to-one) can be improved by $13.53\%$.
 \textbf{2) From the perspective of poisoning ratios}, LPS strategy can be widely applied to different poisoning ratios, but the degree of improvement is also related to the poisoning ratio. 
Specifically, when the poisoning ratio is extremely low (\eg, 1 Img/Cls, $0.054\%$ pratio), although the improvement of our method  is not obvious compared with other strategies due to the attack itself being weak, it also shows similar results.
However, once  the poisoning ratio is increased, LPS shows a strong advantage over other strategies.
\textbf{3) From the perspective of attacks}, our LPS strategy consistently improves different types of triggers and poisoning labels, demonstrating that LPS strategy is widely applicable to various backdoor attacks.

\textbf{Compare with different datasets.}
To verify whether our proposed LPS strategy supports  larger datasets (more images and classes,  larger image size), we also evaluate these three strategies on CIFAR-100 and Tiny-ImageNet. 
The results in~\cref{tab:cifar100_rn18_rn34,tab:tiny_rn18_rn34} further demonstrate the superiority of LPS strategy to both the random selection and the FUS strategy.

\textbf{Resistance to backdoor defense.}
\input{tables/cifar10_resnet18_defense}
We further evaluate the resistance against defenses of different poisoning sample selection strategies.
The defense results are shown in \cref{tab:defense}. 
It can be seen our method outperforms others in most cases (higher ASR is better),  indicating that a reasonable poisoning sample selection strategy probably makes the attack better resistant to defenses.

\subsection{Ablation studies}
\label{sec:ablation}

\begin{wrapfigure}{r}{0.5\linewidth}
\vspace{-5mm}
\centering 
\captionof{table}{ Ablation studies of LPS’s constraints.}
\vspace{-1mm}
\label{tab:ablation}
\renewcommand\arraystretch{1.1}
\resizebox{\linewidth}{!}{
\begin{tabular}{lccccc}
\toprule
Attack  & Pratio & LPS           & LPS$\backslash_{\text{ET}}$ & LPS$\backslash_{\text{ET,PC}}$& FUS\cite{ijcai2022forget} \\ \midrule
BadNets\cite{badnet}  & 0.216\%      & {80.58} & 75.33   & 71.47      & 68.01  \\
Blended\cite{blended}  & 0.432\%      & {87.20} & 85.72   & 82.71      & 79.06  \\ 
SSBA\cite{ssba}     & 0.432\%      & {23.29} & 21.18   & 20.36      & 14.86  \\
Trojan-WM\cite{trojannn}  & 0.216\%    & {93.27} & 89.91   & 87.80      & 77.63  \\
\bottomrule
\end{tabular}}
\vspace{-2mm}
\end{wrapfigure}
\textbf{Effects of different constraints in LPS.} 
As demonstrated under \cref{eq:minmax}, the equation $\mH\vm= \tilde{\alpha}  \cdot \bm\mu$ captures three constraints, including satisfying the poisoning ratio, excluding the target class (dubbed ET), and selecting the same number of samples per class (dubbed PC), respectively. Here we compare LPS with its two variants of changing the last two constraints, including: 
\textbf{1)} \textit{LPS without excluding target class} (LPS$\backslash_{\text{ET}}$),
\textbf{2)} \textit{LPS$\backslash_{\text{ET}}$ without selecting the same number of poisoned samples per class} (LPS$\backslash_{\text{ET,PC}}$).
The results in \cref{tab:ablation} show that both constraints are important for the LPS strategy. 
Note that even removing two constraints, LPS$\backslash_{\text{ET,PC}}$ still outperforms FUS.

\begin{wrapfigure}{r}{0.5\linewidth}
\vspace{-5mm}
\centering 
\includegraphics[width=\linewidth]{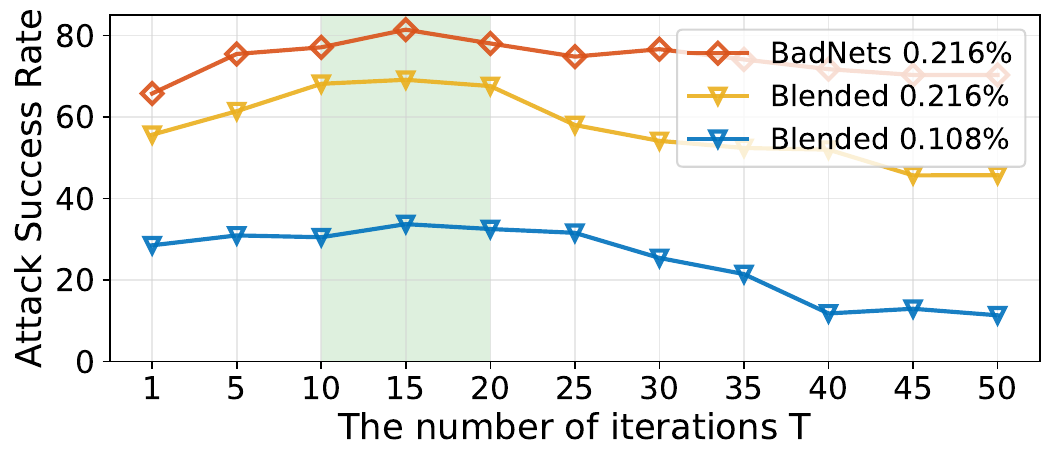}
\vspace{-7mm}
\caption{Attack results of LPS strategy on CIFAR-10 under different iterations $T$.} 
\label{fig:search_step} 
\end{wrapfigure}
\textbf{Effect of the number of iterations $T$.} In \cref{alg:algorithm}, our LPS method requires iteratively solving a min-max optimization problem. 
Here we explore the effect of different iterations $T$ on the attack results.
As shown in \cref{fig:search_step}, we evaluate LPS strategy in a wide range of iterations from $1$ to $50$. 
We can see that LPS strategy shows stable and high performance in the range $T\in[10,20]$. 
Therefore, we choose $T=15$ as the default setting of the main experiments.

%% file: tables/cifar10_rn18_rn34.tex
\begin{table}[t]
\centering
\caption{Attack success rate (\%) on CIFAR-10, where the surrogate and target model are ResNet-18 and ResNet-34 respectively. \textbf{Bold} means the best.}
\label{tab:cifar10_rn18_rn34}
\resizebox{0.9\textwidth}{!}{%
\begin{tabular}{@{}cllllll@{}}
\toprule
\multicolumn{7}{c}{\textbf{Dataset: CIFAR-10\quad Surrogate: ResNet-18   $\Longrightarrow$ Target: ResNet-34}} \\ \midrule
Attack &
  Pratio  (\#Img/Cls) &
  0.054\% (\#3)             &
  0.108\% (\#6)               &
  0.216\% (\#12)              &
  0.432\% (\#24)              &
  0.864\% (\#48)         \\ \midrule
\multirow{3}{*}{\begin{tabular}[c]{@{}c@{}}BadNets~\cite{badnet}\\      (all-to-one)\end{tabular}} &
  Random &
  \textbf{0.86} \footnotesize{ $\pm$ 0.09} &
  1.71 \footnotesize{ $\pm$ 0.48} &
  62.57 \footnotesize{ $\pm$ 5.15} &
  81.71 \footnotesize{ $\pm$ 1.51} &
  89.21 \footnotesize{ $\pm$ 1.05} \\
 &
  FUS~\cite{forgetting} &
  0.75 \footnotesize{ $\pm$ 0.08} &
  1.37 \footnotesize{ $\pm$ 0.22} &
  64.67 \footnotesize{ $\pm$ 5.88} &
  83.41 \footnotesize{ $\pm$ 2.09} &
  90.05 \footnotesize{ $\pm$ 0.34} \\
 &
  LPS (Ours) &
  0.77 \footnotesize{ $\pm$ 0.04} &
  \textbf{5.70} \footnotesize{ $\pm$ 1.77} &
  \textbf{76.41} \footnotesize{ $\pm$ 5.03} &
  \textbf{85.77} \footnotesize{ $\pm$ 5.43} &
  \textbf{91.62} \footnotesize{ $\pm$ 1.25} \\ \midrule
\multirow{3}{*}{\begin{tabular}[c]{@{}c@{}}BadNets~\cite{badnet}\\      (all-to-all)\end{tabular}} &
  Random &
  0.69 \footnotesize{ $\pm$   0.05} &
  0.73 \footnotesize{ $\pm$   0.06} &
  0.93 \footnotesize{ $\pm$   0.22} &
  39.91 \footnotesize{ $\pm$   14.35} &
  75.54 \footnotesize{ $\pm$   2.84} \\
 &
  FUS~\cite{forgetting} &
  \textbf{0.72} \footnotesize{ $\pm$ 0.01} &
  0.75\footnotesize{ $\pm$ 0.02} &
  1.03 \footnotesize{ $\pm$ 0.13} &
  33.37 \footnotesize{ $\pm$ 2.60} &
  76.76 \footnotesize{ $\pm$ 0.24} \\
 &
  LPS (Ours) &
  0.70 \footnotesize{ $\pm$ 0.10} &
  \textbf{0.76 }\footnotesize{ $\pm$ 0.05} &
  \textbf{36.64} \footnotesize{ $\pm$ 2.71} &
  \textbf{66.95}  \footnotesize{ $\pm$ 1.02}&
  \textbf{80.18} \footnotesize{ $\pm$ 2.08} \\ \midrule
\multirow{3}{*}{\begin{tabular}[c]{@{}c@{}}Blended~\cite{blended}\\      (all-to-one)\end{tabular}} &
  Random &
  8.87 \footnotesize{ $\pm$   2.75} &
  23.69 \footnotesize{ $\pm$   3.09} &
  50.65 \footnotesize{ $\pm$   4.05} &
  75.67 \footnotesize{ $\pm$   1.51} &
  89.47 \footnotesize{ $\pm$   0.93} \\
 &
  FUS~\cite{forgetting} &
  \textbf{10.51} \footnotesize{ $\pm$ 2.01} &
  22.29 \footnotesize{ $\pm$ 0.81} &
  51.13 \footnotesize{ $\pm$ 2.83} &
  80.46 \footnotesize{ $\pm$ 1.18} &
  92.11 \footnotesize{ $\pm$ 0.79} \\
 &
  LPS (Ours) &
  9.09 \footnotesize{ $\pm$ 3.54} &
  \textbf{29.84} \footnotesize{ $\pm$ 3.86} &
  \textbf{64.6} \footnotesize{ $\pm$ 4.51}&
  \textbf{87.16} \footnotesize{ $\pm$ 0.84} &
  \textbf{97.53} \footnotesize{ $\pm$ 0.19} \\ \midrule
\multirow{3}{*}{\begin{tabular}[c]{@{}c@{}}Blended~\cite{blended}\\      (all-to-all)\end{tabular}} &
  Random &
  2.48 \footnotesize{ $\pm$   0.11} &
  4.05 \footnotesize{ $\pm$   0.66} &
  9.32 \footnotesize{ $\pm$   1.25} &
  37.33 \footnotesize{ $\pm$   5.01} &
  67.54 \footnotesize{ $\pm$   1.12} \\
 &
  FUS~\cite{forgetting} &
  2.40 \footnotesize{ $\pm$ 0.16} &
  4.17 \footnotesize{ $\pm$ 0.60} &
  6.67 \footnotesize{ $\pm$ 0.49} &
  29.54 \footnotesize{ $\pm$ 2.34} &
  64.90 \footnotesize{ $\pm$ 2.02} \\
 &
  LPS (Ours) &
  \textbf{3.35} \footnotesize{ $\pm$ 0.45} &
  \textbf{7.37} \footnotesize{ $\pm$ 0.78} &
  \textbf{34.6} \footnotesize{ $\pm$ 2.34} &
  \textbf{60.12} \footnotesize{ $\pm$ 1.60} &
  \textbf{72.92} \footnotesize{ $\pm$ 1.00} \\ \midrule
\multirow{3}{*}{\begin{tabular}[c]{@{}c@{}}SIG~\cite{sig}\\      (clean label)\end{tabular}} &
  Random &
  3.48 \footnotesize{ $\pm$ 0.74} &
  6.16 \footnotesize{ $\pm$ 1.74} &
  11.98 \footnotesize{ $\pm$ 0.75} &
  18.72 \footnotesize{ $\pm$ 3.18} &
  36.46 \footnotesize{ $\pm$ 5.34} \\
 &
  FUS~\cite{forgetting} &
  3.30 \footnotesize{ $\pm$ 0.59} &
  8.67 \footnotesize{ $\pm$ 2.10} &
  16.06 \footnotesize{ $\pm$ 3.16} &
  28.50 \footnotesize{ $\pm$ 1.14} &
  46.99 \footnotesize{ $\pm$ 8.77} \\
 &
  LPS (Ours) &
  \textbf{11.38} \footnotesize{ $\pm$ 1.50} &
  \textbf{19.00} \footnotesize{ $\pm$ 1.66} &
  \textbf{32.67} \footnotesize{ $\pm$ 3.06} &
  \textbf{51.32} \footnotesize{ $\pm$ 4.17} &
  \textbf{65.77} \footnotesize{ $\pm$ 5.80} \\ \midrule
\multirow{3}{*}{\begin{tabular}[c]{@{}c@{}}SSBA~\cite{ssba}\\      (all-to-one)\end{tabular}} &
  Random &
  1.01 \footnotesize{ $\pm$   0.12} &
  \textbf{1.05} \footnotesize{ $\pm$   0.04} &
  2.06 \footnotesize{ $\pm$   0.15} &
  20.34 \footnotesize{ $\pm$   5.58} &
  60.36 \footnotesize{ $\pm$   2.42} \\
 &
  FUS~\cite{forgetting} &
  \textbf{1.10} \footnotesize{ $\pm$ 0.16} &
  1.04 \footnotesize{ $\pm$ 0.28} &
  2.02 \footnotesize{ $\pm$ 0.45} &
  16.81 \footnotesize{ $\pm$ 3.47} &
  60.64 \footnotesize{ $\pm$ 3.29} \\
 &
  LPS (Ours) &
  0.98 \footnotesize{ $\pm$ 0.17} &
  1.03 \footnotesize{ $\pm$ 0.05} &
  \textbf{2.30 }\footnotesize{ $\pm$ 0.51} &
  \textbf{22.92 }\footnotesize{ $\pm$ 2.74} &
  \textbf{64.39 }\footnotesize{ $\pm$ 2.96} \\ \midrule
\multirow{3}{*}{\begin{tabular}[c]{@{}c@{}}Trojan-WM~\cite{trojannn}\\      (all-to-one)\end{tabular}} &
  Random &
  3.39 \footnotesize{ $\pm$ 1.37} &
  23.26 \footnotesize{ $\pm$ 11.74} &
  80.04 \footnotesize{ $\pm$ 7.19} &
  94.96 \footnotesize{ $\pm$ 1.92} &
  98.27 \footnotesize{ $\pm$ 0.34} \\
 &
  FUS~\cite{forgetting} &
  3.07 \footnotesize{ $\pm$ 1.62} &
  19.22 \footnotesize{ $\pm$ 6.12} &
  78.85 \footnotesize{ $\pm$ 4.70} &
  96.59 \footnotesize{ $\pm$ 1.57} &
  99.25 \footnotesize{ $\pm$ 0.38} \\
 &
  LPS (Ours) &
  \textbf{3.66 }\footnotesize{ $\pm$ 0.33} &
  \textbf{33.77} \footnotesize{ $\pm$ 10.47} &
  \textbf{94.32} \footnotesize{ $\pm$ 0.81} &
  \textbf{99.77} \footnotesize{ $\pm$ 0.06} &
  \textbf{99.97} \footnotesize{ $\pm$ 0.01} \\ \bottomrule
\end{tabular}%
}
\vspace{-6mm}
\end{table}

%% file: tables/cifar100_rn18_rn34.tex
\begin{table}[t]
\centering
\caption{Attack success rate (\%) on CIFAR-100, where the surrogate and target model are ResNet-18 and ResNet-34 respectively. \textbf{Bold} means the best.}
\label{tab:cifar100_rn18_rn34}
\resizebox{0.9\textwidth}{!}{%
\begin{tabular}{@{}cllllll@{}}
\toprule
\multicolumn{7}{c}{\textbf{Dataset: CIFAR-100 \quad Surrogate: ResNet-18   $\Longrightarrow$ Target: ResNet-34}} \\ \midrule
\multicolumn{1}{l}{Attack} &
  Pratio  (\#Img/Cls)&
  0.198\% (\#1)      &
  0.396\% (\#2)      &
  0.594\% (\#3)     &
  0.792\% (\#4)     &
  0.99 \% (\#5)       \\ \midrule
\multirow{3}{*}{\begin{tabular}[c]{@{}c@{}}BadNets~\cite{badnet}\\      (all-to-one)\end{tabular}} &
  Random &
  8.09 \footnotesize{ $\pm$    2.31} &
  36.74 \footnotesize{ $\pm$    6.22} &
  50.68 \footnotesize{ $\pm$    2.68} &
  59.50 \footnotesize{ $\pm$    4.56} &
  64.81 \footnotesize{ $\pm$    5.97} \\
 &
  FUS~\cite{forgetting} &
  10.41 \footnotesize{ $\pm$  4.20} &
  43.60 \footnotesize{ $\pm$  6.79} &
  51.06 \footnotesize{ $\pm$  6.74} &
  62.28 \footnotesize{ $\pm$  6.22} &
  68.34 \footnotesize{ $\pm$  6.30} \\
 &
  LPS (Ours) &
  \textbf{17.98} \footnotesize{ $\pm$  2.58} &
  \textbf{52.02} \footnotesize{ $\pm$  4.05} &
  \textbf{58.46} \footnotesize{ $\pm$  1.87} &
  \textbf{63.49} \footnotesize{ $\pm$  5.90} &
  \textbf{70.45} \footnotesize{ $\pm$  3.49} \\ \midrule
\multirow{3}{*}{\begin{tabular}[c]{@{}c@{}}Blended~\cite{blended}\\      (all-to-one)\end{tabular}} &
  Random &
  37.53 \footnotesize{ $\pm$  2.23} &
  59.98 \footnotesize{ $\pm$  1.09} &
  68.53 \footnotesize{ $\pm$  1.26} &
  77.37 \footnotesize{ $\pm$  0.67} &
  81.47 \footnotesize{ $\pm$  0.26} \\
 &
  FUS~\cite{forgetting} &
  \textbf{38.65} \footnotesize{ $\pm$  1.58} &
  65.75 \footnotesize{ $\pm$  0.98} &
  69.04 \footnotesize{ $\pm$  5.40} &
  82.25 \footnotesize{ $\pm$  0.69} &
  86.14 \footnotesize{ $\pm$  0.46} \\
 &
  LPS (Ours) &
  38.64 \footnotesize{ $\pm$  1.72} &
  \textbf{66.94} \footnotesize{ $\pm$  1.75} &
  \textbf{81.73} \footnotesize{ $\pm$  1.73} &
  \textbf{89.88} \footnotesize{ $\pm$  1.19} &
  \textbf{93.29} \footnotesize{ $\pm$  0.67} \\ \midrule
\multirow{3}{*}{\begin{tabular}[c]{@{}c@{}}SIG~\cite{sig}\\      (clean label)\end{tabular}} &
  Random &
  2.79 \footnotesize{ $\pm$  0.44} &
  6.09 \footnotesize{ $\pm$  0.99} &
  14.3 \footnotesize{ $\pm$  2.38} &
  22.08 \footnotesize{ $\pm$  3.28} &
  43.95 \footnotesize{ $\pm$  1.35} \\
 &
  FUS~\cite{forgetting} &
  3.79 \footnotesize{ $\pm$  1.12} &
  \textbf{7.80} \footnotesize{ $\pm$  1.60} &
  15.84 \footnotesize{ $\pm$  2.50} &
  N/A &
  N/A \\
 &
  LPS (Ours) &
  \textbf{4.49} \footnotesize{ $\pm$  1.43} &
  7.01 \footnotesize{ $\pm$    1.73} &
  \textbf{16.11} \footnotesize{ $\pm$  1.99} &
  \textbf{25.12} \footnotesize{ $\pm$    1.37} &
  \textbf{46.43} \footnotesize{ $\pm$    0.55} \\ \midrule
\multirow{3}{*}{\begin{tabular}[c]{@{}c@{}}SSBA~\cite{ssba}\\      (all-to-one)\end{tabular}} &
  Random &
  1.42 \footnotesize{ $\pm$  0.24} &
  7.45 \footnotesize{ $\pm$    1.62} &
  18.73 \footnotesize{ $\pm$  3.33} &
  31.61 \footnotesize{ $\pm$    0.63} &
  43.37 \footnotesize{ $\pm$    1.77} \\
 &
  FUS~\cite{forgetting} &
  \textbf{1.51} \footnotesize{ $\pm$  0.40} &
  7.99 \footnotesize{ $\pm$  1.11} &
  18.44 \footnotesize{ $\pm$  1.51} &
  33.35 \footnotesize{ $\pm$  1.06} &
  44.00 \footnotesize{ $\pm$  2.66} \\
 &
  LPS (Ours) &
  1.49 \footnotesize{ $\pm$  0.10} &
  \textbf{8.03} \footnotesize{ $\pm$  1.09} &
  \textbf{21.46} \footnotesize{ $\pm$  1.81} &
  \textbf{34.12} \footnotesize{ $\pm$  2.85} &
  \textbf{48.77} \footnotesize{ $\pm$  3.18} \\ \midrule
\multirow{3}{*}{\begin{tabular}[c]{@{}c@{}}Trojan-WM~\cite{trojannn}\\      (all-to-one)\end{tabular}} &
  Random &
  39.44 \footnotesize{ $\pm$    4.24} &
  68.64 \footnotesize{ $\pm$    1.83} &
  82.13 \footnotesize{ $\pm$    0.47} &
  88.08 \footnotesize{ $\pm$    0.93} &
  91.16 \footnotesize{ $\pm$    1.52} \\
 &
  FUS~\cite{forgetting} &
  39.74 \footnotesize{ $\pm$  2.42} &
  75.43 \footnotesize{ $\pm$  3.23} &
  84.80 \footnotesize{ $\pm$  0.79} &
  92.58 \footnotesize{ $\pm$  0.95} &
  93.87 \footnotesize{ $\pm$  0.33} \\
 &
  LPS (Ours) &
  \textbf{44.90} \footnotesize{ $\pm$  3.51} &
  \textbf{84.75} \footnotesize{ $\pm$  3.24} &
  \textbf{96.36} \footnotesize{ $\pm$  1.18} &
  \textbf{98.16} \footnotesize{ $\pm$  0.33} &
  \textbf{99.30} \footnotesize{ $\pm$  0.16} \\ \bottomrule
\end{tabular}%

}
\vspace{-5mm}
\end{table}

%% file: tables/tiny_rn18_rn34.tex
\begin{table}[t]
\centering
\caption{Attack success rate (\%) on Tiny-ImageNet, where the surrogate and target model are ResNet-18 and ResNet-34 respectively. \textbf{Bold} means the best.}
\label{tab:tiny_rn18_rn34}
\resizebox{0.9\textwidth}{!}{%
\begin{tabular}{@{}cllllll@{}}
\toprule
\multicolumn{7}{c}{\textbf{Dataset: Tiny-ImageNet \quad Surrogate: ResNet-18   $\Longrightarrow$ Target: ResNet-34}} \\ \midrule
\multicolumn{1}{l}{Attack} &
  Pratio (\#Img/Cls) &
  0.199\% (\#1) &
  0.398\% (\#2) &
  0.597\% (\#3) &
  0.796\% (\#4) &
  0.995\% (\#5) \\ \midrule
\multirow{3}{*}{\begin{tabular}[c]{@{}c@{}}BadNets~\cite{badnet}\\      (all-to-one)\end{tabular}} &
  Random &
  4.93 \footnotesize{ $\pm$ 6.19} &
  37.18 \footnotesize{ $\pm$ 6.61} &
  42.98 \footnotesize{ $\pm$ 1.89} &
  48.91 \footnotesize{ $\pm$ 3.46} &
  60.52 \footnotesize{ $\pm$ 2.35} \\
 &
  FUS~\cite{forgetting} &
  \textbf{5.44}  \footnotesize{ $\pm$ 3.54}&
  32.93 \footnotesize{ $\pm$ 1.69} &
  43.74 \footnotesize{ $\pm$ 3.67} &
  48.72 \footnotesize{ $\pm$ 3.58} &
  60.76 \footnotesize{ $\pm$ 4.72} \\
 &
  LPS (Ours) &
  5.21 \footnotesize{ $\pm$ 3.10} &
  \textbf{38.05} \footnotesize{ $\pm$ 2.26} &
  \textbf{47.21} \footnotesize{ $\pm$ 3.90} &
  \textbf{49.34} \footnotesize{ $\pm$ 3.41} &
  \textbf{61.22} \footnotesize{ $\pm$ 2.12} \\ \midrule
\multirow{3}{*}{\begin{tabular}[c]{@{}c@{}}Blended~\cite{blended}\\      (all-to-one)\end{tabular}} &
  Random &
  66.73 \footnotesize{ $\pm$ 0.52} &
  78.79 \footnotesize{ $\pm$ 0.63} &
  84.87 \footnotesize{ $\pm$ 1.50} &
  87.81 \footnotesize{ $\pm$ 0.72} &
  89.96 \footnotesize{ $\pm$ 0.43} \\
 &
  FUS~\cite{forgetting} &
  70.95 \footnotesize{ $\pm$ 1.47} &
  82.01 \footnotesize{ $\pm$ 0.50} &
  88.38 \footnotesize{ $\pm$ 0.94} &
  90.70 \footnotesize{ $\pm$ 1.37} &
  93.19 \footnotesize{ $\pm$ 0.39} \\
 &
  LPS (Ours) &
  \textbf{82.76} \footnotesize{ $\pm$ 2.52} &
  \textbf{93.55} \footnotesize{ $\pm$ 0.45} &
  \textbf{96.20} \footnotesize{ $\pm$ 0.11} &
  \textbf{97.65} \footnotesize{ $\pm$ 0.10} &
  \textbf{98.08} \footnotesize{ $\pm$ 0.09} \\ \midrule
\multirow{3}{*}{\begin{tabular}[c]{@{}c@{}}SIG\cite{sig}\\      (all-to-one)\end{tabular}} &
  Random &
  61.80 \footnotesize{ $\pm$ 3.30} &
  81.15 \footnotesize{ $\pm$ 0.62} &
  87.87 \footnotesize{ $\pm$ 1.83} &
  90.80 \footnotesize{ $\pm$ 0.55} &
  92.77 \footnotesize{ $\pm$ 0.95} \\
 &
  FUS~\cite{forgetting} &
  60.02 \footnotesize{ $\pm$ 1.76} &
  84.95 \footnotesize{ $\pm$ 2.53} &
  86.36 \footnotesize{ $\pm$ 6.11} &
  92.47 \footnotesize{ $\pm$ 1.41} &
  94.56 \footnotesize{ $\pm$ 0.59} \\
 &
  LPS (Ours) &
  \textbf{62.90} \footnotesize{ $\pm$ 3.07} &
  \textbf{91.57} \footnotesize{ $\pm$ 2.00} &
  \textbf{96.59} \footnotesize{ $\pm$ 1.07} &
  \textbf{98.02} \footnotesize{ $\pm$ 0.45} &
  \textbf{98.97} \footnotesize{ $\pm$ 0.17} \\ \midrule
\multirow{3}{*}{\begin{tabular}[c]{@{}c@{}}SSBA~\cite{ssba}\\      (all-to-one)\end{tabular}} &
  Random &
  34.34 \footnotesize{ $\pm$ 2.93} &
  60.05 \footnotesize{ $\pm$ 3.32} &
  76.09 \footnotesize{ $\pm$ 0.88} &
  81.60 \footnotesize{ $\pm$ 0.25} &
  85.65 \footnotesize{ $\pm$ 0.30} \\
 &
  FUS~\cite{forgetting} &
  \textbf{34.80} \footnotesize{ $\pm$ 0.71} &
  60.68 \footnotesize{ $\pm$ 1.58} &
  76.83 \footnotesize{ $\pm$ 1.55} &
  84.53 \footnotesize{ $\pm$ 0.54} &
  88.48 \footnotesize{ $\pm$ 0.51} \\
 &
  LPS (Ours) &
  33.68 \footnotesize{ $\pm$ 0.96} &
  \textbf{61.58} \footnotesize{ $\pm$ 1.27} &
  \textbf{82.61} \footnotesize{ $\pm$ 0.39} &
  \textbf{91.72} \footnotesize{ $\pm$ 2.53} &
  \textbf{94.5 \footnotesize{ $\pm$ 0.80}} \\ \midrule
\multirow{3}{*}{\begin{tabular}[c]{@{}c@{}}Trojan-WM~\cite{trojannn}\\      (all-to-one)\end{tabular}} &
  Random &
  \textbf{6.75} \footnotesize{ $\pm$ 1.31} &
  28.55 \footnotesize{ $\pm$ 3.17} &
  61.06 \footnotesize{ $\pm$ 3.9} &
  74.18 \footnotesize{ $\pm$ 0.42} &
  80.74 \footnotesize{ $\pm$ 0.77} \\
 &
  FUS~\cite{forgetting} &
  6.35 \footnotesize{ $\pm$ 1.13} &
  26.47 \footnotesize{ $\pm$ 6.56} &
  51.05 \footnotesize{ $\pm$ 6.86} &
  75.21 \footnotesize{ $\pm$ 3.15} &
  85.34 \footnotesize{ $\pm$ 1.20} \\
 &
  LPS (Ours) &
 6.26 \footnotesize{ $\pm$ 0.81} &
  \textbf{48.26} \footnotesize{ $\pm$ 8.35} &
  \textbf{77.56} \footnotesize{ $\pm$ 2.93} &
  \textbf{86.16} \footnotesize{ $\pm$ 1.85} &
  \textbf{92.89} \footnotesize{ $\pm$ 1.72} \\ \bottomrule
\end{tabular}%
}
\vspace{-4mm}
\end{table}

%% file: tables/cifar10_resnet18_defense.tex
\begin{table}[]
\centering
\caption{Results of various defenses against attacks on CIFAR-10. \textbf{Bold} means the best}
\label{tab:defense}
\resizebox{0.95\textwidth}{!}{%
\begin{tabular}{@{}c|l|ll|ll|ll|ll|ll|ll|ll@{}}
\toprule
\multirow{2}{*}{Attack} &
  \multicolumn{1}{c|}{\multirow{2}{*}{Defense}} &
  \multicolumn{2}{c|}{No Defense} &
  \multicolumn{2}{c|}{FT} &
  \multicolumn{2}{c|}{FP\cite{FP}} &
  \multicolumn{2}{c|}{ABL\cite{abl}} &
  \multicolumn{2}{c|}{NAD\cite{nad}} &
  \multicolumn{2}{c|}{CLP\cite{clp}} &
  \multicolumn{2}{c}{I-BAU\cite{ibau}} \\ 
 &
  \multicolumn{1}{c|}{} &
  ASR &
  ACC &
  ASR &
  ACC &
  ASR &
  ACC &
  ASR &
  ACC &
  ASR &
  ACC &
  ASR &
  ACC &
  ASR &
  ACC \\ \midrule
\multirow{3}{*}{\begin{tabular}[c]{@{}c@{}}BadNets\cite{badnet} \\0.216\%\\      (all-to-one)\end{tabular}} &
  Random &
  69.73 &
  93.97 &
  35.82 &
  93.87 &
  3.88 &
  93.43 &
  18.86 &
  62.42 &
  1.18 &
  88.23 &
  9.57 &
  92.91 &
  2.18 &
  84.64 \\
 &
  FUS\cite{ijcai2022forget} &
  68.97 &
  93.74 &
  39.31 &
  93.99 &
  6.3 &
  93.59 &
  18.84 &
  72.94 &
  1.82 &
  87.34 &
  35.51 &
  93.65 &
  3.86 &
  76.67 \\
 &
  LPS (Ours) &
  \textbf{81.94} &
  93.76 &
  \textbf{51.48} &
  92.57 &
  \textbf{10.88} &
  93.45 &
  \textbf{23.17} &
  63.59 &
  \textbf{7.33} &
  91.77 &
  \textbf{39.29} &
  93.35 &
  \textbf{7.01} &
  88.99 \\ \midrule
\multirow{3}{*}{\begin{tabular}[c]{@{}c@{}}Blended\cite{blended}  \\ 0.216\%\\      (all-to-one)\end{tabular}} &
  Random &
  53.22 &
  94.01 &
  33.26 &
  93.85 &
  24.39 &
  93.47 &
  30.07 &
  71.75 &
  23.58 &
  91.59 &
  32.53 &
  93.33 &
  9.77 &
  76.32 \\
 &
  FUS\cite{ijcai2022forget} &
  48.96 &
  93.99 &
  34.04 &
  93.94 &
  21.67 &
  93.54 &
  29.19 &
  75.84 &
  25.16 &
  92.83 &
  \textbf{38.51} &
  93.62 &
  6.29 &
  83.6 \\
 &
  LPS (Ours) &
 \textbf{ 59.73} &
  93.96 &
  \textbf{34.68} &
  93.23 &
  \textbf{28.02} &
  93.79 &
  \textbf{38.01} &
  71.92 &
  \textbf{25.98} &
  91.56 &
  37.66 &
  93.37 &
  \textbf{9.18} &
  75.7 \\ \midrule
\multirow{3}{*}{\begin{tabular}[c]{@{}c@{}}SIG\cite{sig} \\0.216\%\\      (clean label)\end{tabular}} &
  Random &
  12.61 &
  93.86 &
  12.58 &
  93.59 &
  10.84 &
  93.45 &
  13.99 &
  73.69 &
  2.08 &
  90.88 &
  15.48 &
  93.63 &
  2.99 &
  87.26 \\
 &
  FUS\cite{ijcai2022forget} &
  14.19 &
  93.88 &
  11.83 &
  93.87 &
  12.81 &
  93.44 &
  10.91 &
  76.7 &
  4.21 &
  90.34 &
  15.04 &
  93.27 &
  6.31 &
  84.96 \\
 &
  LPS (Ours) &
 \textbf{ 41.31 }&
  93.82 &
  \textbf{38.01} &
  93.94 &
  \textbf{36.59} &
  93.52 &
  \textbf{34.06 }&
  72.19 &
  \textbf{29.52} &
  91.37 &
  \textbf{48.73 }&
  93.64 &
  \textbf{7.92} &
  89.42 \\ \midrule
\multirow{3}{*}{\begin{tabular}[c]{@{}c@{}}Trojan-WM\cite{trojannn} \\0.216\%\\      (all-to-one)\end{tabular}} &
  Random &
  89.43 &
  93.73 &
  86.4 &
  93.6 &
  \textbf{46.59} &
  93.15 &
  51.71 &
  71.5 &
  43.21 &
  91.2 &
  2.74 &
  92.75 &
  7.29 &
  84.57 \\
 &
  FUS\cite{ijcai2022forget} &
  82.9 &
  93.83 &
  68.7 &
  93.73 &
  35.52 &
  93.57 &
  48.86 &
  74.97 &
  40.48 &
  92.69 &
  6.72 &
  93.41 &
  \textbf{11.96} &
  81.53 \\
 &
  LPS (Ours) &
  \textbf{93.76} &
  94.01 &
  \textbf{86.94} &
  94.17 &
  30.09 &
  93.44 &
  \textbf{62.66} &
  69.58 &
  \textbf{46.65} &
  91.69 &
  \textbf{59.21} &
  93.84 &
  9.70 &
  86.36 \\ \bottomrule
\end{tabular}%
}
\vspace{-4mm}
\end{table}

%% file: 5_analysis.tex
\section{Analysis}
\label{sec:analysis}

\textbf{Analysis of computational complexity.} 
Both LPS and FUS adopt the iterative algorithm by alternatively updating the surrogate model and the poisoning mask. In term of updating the surrogate model in each iteration, the complexity is $O(|\gD|K(F+B))$, with $|\gD|$ being the train data size, $F$ is the cost of forward pass in a DNN model and $B$ being the backward \cite{BP} pass cost, $K$ being the number of epochs. 
In terms of updating the poisoning mask, it requires one forward pass for all training samples, then the complexity is $O(|\gD| F)$. Thus, the overall complexity of both LPS and FUS is $O(T|\gD|((K+1)F+KB))$, $T$ being the number of overall iterations.  
It is notable that in FUS, the surrogate model is re-initialized in each iteration, so it has to set $K$ as a large number (\ie, 60), while our LPS sets $K$ as 1. Thus, our practical efficiency is much better than FUS. We compare the full training time of different strategies in the \textbf{supplement materials}.

\begin{wrapfigure}{r}{0.65\linewidth}
\vspace{-5mm}
\centering 
\includegraphics[width=\linewidth]{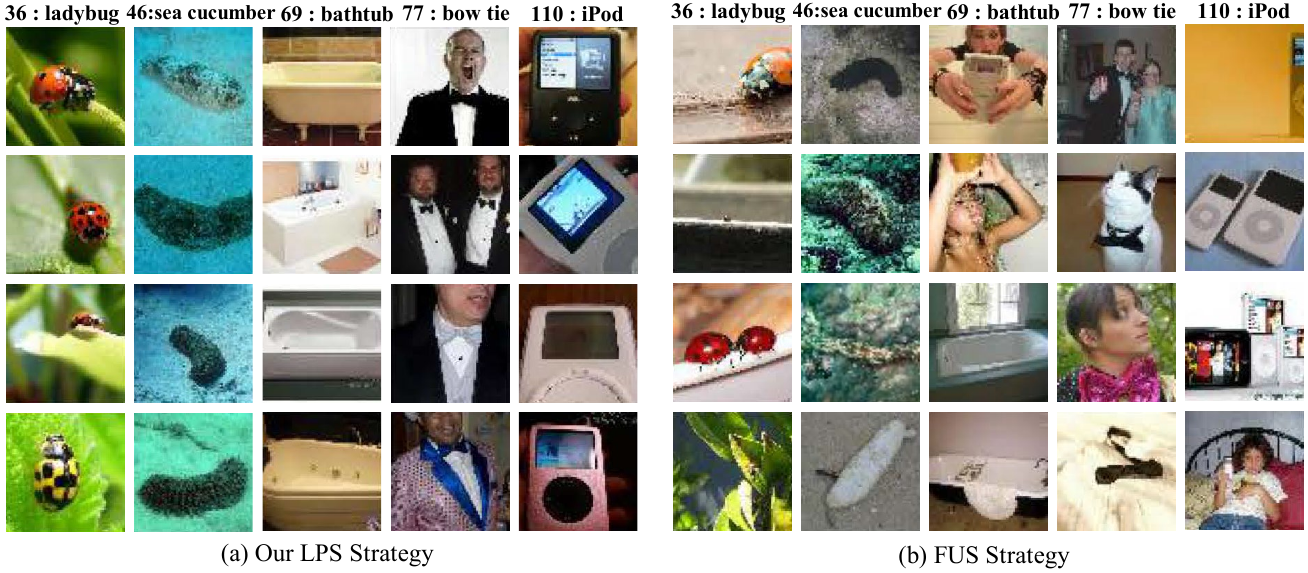}
\vspace{-5mm}
\caption{Visualization of samples selected by our LPS (a) and FUS (b).} 
\label{fig:visualization} 
\vspace{-3mm}
\end{wrapfigure}
\begin{figure} 
\begin{minipage}[t]{0.47\linewidth} 
\centering 
\includegraphics[width=\linewidth]{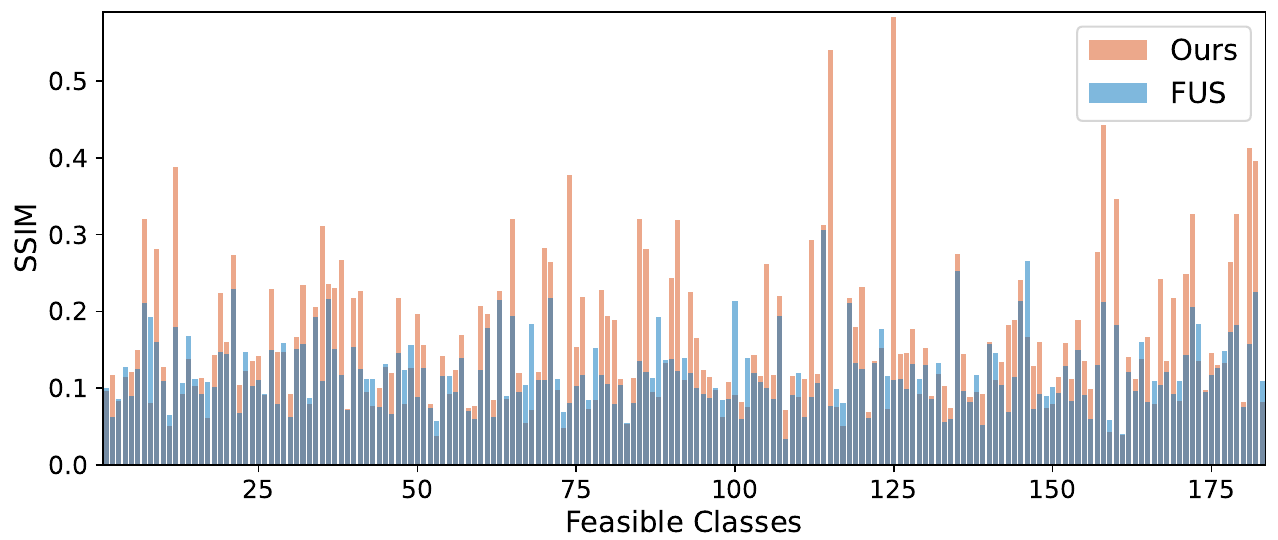}
\caption{Average pairwise SSIM for each class computed over samples selected by our method and FUS on Tiny-ImageNet.} 
\label{fig:ssim} 
\end{minipage}\hspace{1mm}
\begin{minipage}[t]{0.47\linewidth} 
\centering 
\includegraphics[width=\linewidth]{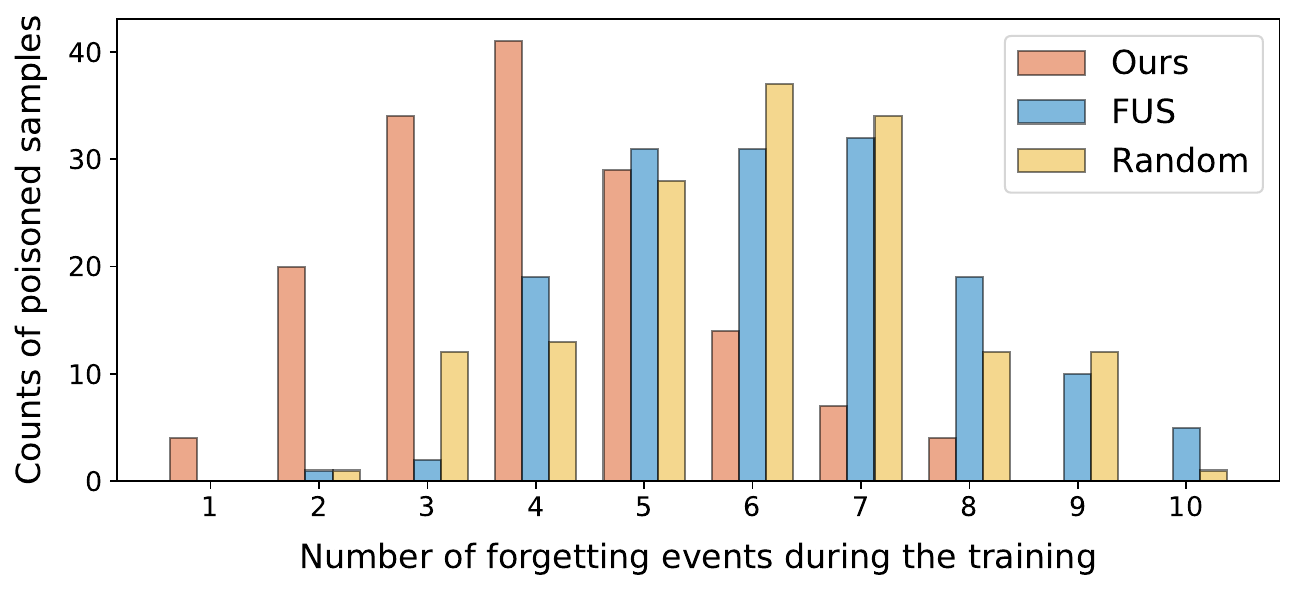}
\caption{The distribution of the forgetting events of poisoned samples on CIFAR-10. We use Blended attack the poisoning ratio is $0.306\%$} 
\label{fig:forget}
\end{minipage} 
\vspace{-6mm}
\end{figure}

\textbf{Visualization of selected samples.}
In  \cref{fig:visualization}, we visualize some samples selected by our method and FUS from Tiny-ImageNet, from which we can find that our method differs from FUS in two aspects. First, our method tends to select samples with discriminative patterns that is easy to remember. It indicates that our method prefers samples with higher clean confidence. Second, the samples selected by our method have a higher inter-class similarity. To evaluate the inter-class similarity, we compute the average pairwise Structural Similarity Index (SSIM)\cite{wang2004image} within each class over samples selected by our method and FUS, respectively. Since some classes are ignored by FUS, we only report the classes selected by both our method and FUS. The results are reported in  \cref{fig:ssim} which show that our LPS has higher inter-class similarity.

\textbf{The importance  of selected samples.}
In  \cref{fig:forget}, we present the distribution for forgetting events histogram of blended trigger poisoned samples from CIFAR-10 obtained using different strategies at a very low poisoning ratio. Forgetting events were calculated during the standard training of the target model, given the poisoning masks obtained by different strategies. The results show that DNN trained with poisoned sample whose forgetting events is small have higher generalization performance. DNN do not force a poisoned sample in mind, losing generalization capability.

%% file: 6_conclusion.tex
\section{Conclusion and future work}

This work has explored an often overlooked step in data-poisoning based backdoor attacks, \ie, selecting which benign samples to generate poisoned samples. 
We innovatively propose a learnable poisoning sample selection strategy based on the trigger and benign data. 
It is formulated as a min-max optimization problem, where a surrogate model and a binary poisoning mask are learned together, to encourage the selected samples to have good backdoor effect when training the unknown target model. 
Extensive results validate the effectiveness and efficiency of the proposed LPS strategy in enhancing existing data-poisoning backdoor attacks.

\noindent
\textbf{Limitations and future works}.
Note that in the case of extremely low poisoning ratio, the improvement of LPS is very limited, mainly due to that the poisoning information contained in few poisoned samples with fixed triggers are insufficient to inject backdoor, no matter which poisoning samples are selected. It inspires that learning trigger and poisoning sample selection simultaneously may further enhance the backdoor attack, which will be explored in future. 
In addition, the proposed LPS strategy is specially designed for data poisoning backdoor attack. Developing the similar selection strategy for training controllable backdoor attack also deserves to be explored in future.

\noindent
\textbf{Broader impacts}.
The proposed LPS strategy could be easily utilized by adversaries to enlarge the attack performance of existing backdoor attack methods, which exposes the urgency to develop proactive defense strategies and detection mechanisms to safeguard machine learning systems. %

%% file: main.bbl
\begin{thebibliography}{57}
\providecommand{\natexlab}[1]{#1}
\providecommand{\url}[1]{\texttt{#1}}
\expandafter\ifx\csname urlstyle\endcsname\relax
  \providecommand{\doi}[1]{doi: #1}\else
  \providecommand{\doi}{doi: \begingroup \urlstyle{rm}\Url}\fi

\bibitem[Barni et~al.(2019)Barni, Kallas, and Tondi]{sig}
Mauro Barni, Kassem Kallas, and Benedetta Tondi.
\newblock A new backdoor attack in cnns by training set corruption without
  label poisoning.
\newblock In \emph{2019 IEEE International Conference on Image Processing},
  2019.

\bibitem[Borsos et~al.(2020)Borsos, Mutny, and Krause]{borsos2020coresets}
Zal{\'a}n Borsos, Mojmir Mutny, and Andreas Krause.
\newblock Coresets via bilevel optimization for continual learning and
  streaming.
\newblock \emph{Advances in Neural Information Processing Systems},
  33:\penalty0 14879--14890, 2020.

\bibitem[Bottou and Bousquet(2007)]{SGD}
L{\'e}on Bottou and Olivier Bousquet.
\newblock The tradeoffs of large scale learning.
\newblock \emph{Advances in neural information processing systems}, 20, 2007.

\bibitem[Chai and Chen(2022)]{chai2022oneshot}
Shuwen Chai and Jinghui Chen.
\newblock One-shot neural backdoor erasing via adversarial weight masking.
\newblock In \emph{Thirty-Sixth Conference on Neural Information Processing
  Systems}, 2022.

\bibitem[Chang et~al.(2017)Chang, Learned-Miller, and
  McCallum]{chang2017active}
Haw-Shiuan Chang, Erik Learned-Miller, and Andrew McCallum.
\newblock Active bias: Training more accurate neural networks by emphasizing
  high variance samples.
\newblock \emph{Advances in Neural Information Processing Systems}, 30, 2017.

\bibitem[Chen et~al.(2019)Chen, Carvalho, Baracaldo, Ludwig, Edwards, Lee,
  Molloy, and Srivastava]{ac}
Bryant Chen, Wilka Carvalho, Nathalie Baracaldo, Heiko Ludwig, Benjamin
  Edwards, Taesung Lee, Ian Molloy, and Biplav Srivastava.
\newblock Detecting backdoor attacks on deep neural networks by activation
  clustering.
\newblock In \emph{The AAAI Conference on Artificial Intelligence Workshop},
  2019.

\bibitem[Chen et~al.(2017)Chen, Liu, Li, Lu, and Song]{blended}
Xinyun Chen, Chang Liu, Bo~Li, Kimberly Lu, and Dawn Song.
\newblock Targeted backdoor attacks on deep learning systems using data
  poisoning.
\newblock \emph{arXiv preprint arXiv:1712.05526}, 2017.

\bibitem[Coleman et~al.(2019)Coleman, Yeh, Mussmann, Mirzasoleiman, Bailis,
  Liang, Leskovec, and Zaharia]{coleman2019selection}
Cody Coleman, Christopher Yeh, Stephen Mussmann, Baharan Mirzasoleiman, Peter
  Bailis, Percy Liang, Jure Leskovec, and Matei Zaharia.
\newblock Selection via proxy: Efficient data selection for deep learning.
\newblock \emph{arXiv preprint arXiv:1906.11829}, 2019.

\bibitem[Doan et~al.(2021{\natexlab{a}})Doan, Lao, and Li]{doan2021backdoor}
Khoa Doan, Yingjie Lao, and Ping Li.
\newblock Backdoor attack with imperceptible input and latent modification.
\newblock \emph{Advances in Neural Information Processing Systems},
  34:\penalty0 18944--18957, 2021{\natexlab{a}}.

\bibitem[Doan et~al.(2021{\natexlab{b}})Doan, Lao, Zhao, and Li]{doan2021lira}
Khoa Doan, Yingjie Lao, Weijie Zhao, and Ping Li.
\newblock Lira: Learnable, imperceptible and robust backdoor attacks.
\newblock In \emph{Proceedings of the IEEE/CVF International Conference on
  Computer Vision}, 2021{\natexlab{b}}.

\bibitem[Doan et~al.(2022)Doan, Lao, and Li]{doan2022marksman}
Khoa~D Doan, Yingjie Lao, and Ping Li.
\newblock Marksman backdoor: Backdoor attacks with arbitrary target class.
\newblock In \emph{Thirty-Sixth Conference on Neural Information Processing
  Systems}, 2022.

\bibitem[Doan et~al.(2023)Doan, Lao, and Li]{doan2023bdvits}
Khoa~D Doan, Yingjie Lao, and Ping Li.
\newblock Defending backdoor attacks on vision transformer via patch
  processing.
\newblock In \emph{AAAI Conference on Artificial Intelligence}, 2023.

\bibitem[Gu et~al.(2019)Gu, Liu, Dolan-Gavitt, and Garg]{badnet}
Tianyu Gu, Kang Liu, Brendan Dolan-Gavitt, and Siddharth Garg.
\newblock Badnets: Evaluating backdooring attacks on deep neural networks.
\newblock \emph{IEEE Access}, 7:\penalty0 47230--47244, 2019.

\bibitem[He et~al.(2016)He, Zhang, Ren, and Sun]{resnet}
Kaiming He, Xiangyu Zhang, Shaoqing Ren, and Jian Sun.
\newblock Identity mappings in deep residual networks.
\newblock In \emph{European conference on computer vision}, 2016.

\bibitem[Huang et~al.(2022)Huang, Li, Wu, Qin, and Ren]{huang2022backdoor}
Kunzhe Huang, Yiming Li, Baoyuan Wu, Zhan Qin, and Kui Ren.
\newblock Backdoor defense via decoupling the training process.
\newblock In \emph{International Conference on Learning Representations}, 2022.

\bibitem[Just et~al.(2023)Just, Kang, Wang, Zeng, Ko, Jin, and
  Jia]{just2023lava}
Hoang~Anh Just, Feiyang Kang, Tianhao Wang, Yi~Zeng, Myeongseob Ko, Ming Jin,
  and Ruoxi Jia.
\newblock {LAVA}: Data valuation without pre-specified learning algorithms.
\newblock In \emph{The Eleventh International Conference on Learning
  Representations}, 2023.

\bibitem[Katharopoulos and Fleuret(2018)]{katharopoulos2018not}
Angelos Katharopoulos and Fran{\c{c}}ois Fleuret.
\newblock Not all samples are created equal: Deep learning with importance
  sampling.
\newblock In \emph{International conference on machine learning}, 2018.

\bibitem[Kaushal et~al.(2019)Kaushal, Iyer, Kothawade, Mahadev, Doctor, and
  Ramakrishnan]{kaushal2019learning}
Vishal Kaushal, Rishabh Iyer, Suraj Kothawade, Rohan Mahadev, Khoshrav Doctor,
  and Ganesh Ramakrishnan.
\newblock Learning from less data: A unified data subset selection and active
  learning framework for computer vision.
\newblock In \emph{IEEE Winter Conference on Applications of Computer Vision},
  2019.

\bibitem[Killamsetty et~al.(2021{\natexlab{a}})Killamsetty, Durga,
  Ramakrishnan, De, and Iyer]{killamsetty2021grad}
Krishnateja Killamsetty, S~Durga, Ganesh Ramakrishnan, Abir De, and Rishabh
  Iyer.
\newblock Grad-match: Gradient matching based data subset selection for
  efficient deep model training.
\newblock In \emph{International Conference on Machine Learning},
  2021{\natexlab{a}}.

\bibitem[Killamsetty et~al.(2021{\natexlab{b}})Killamsetty, Sivasubramanian,
  Ramakrishnan, and Iyer]{killamsetty2021glister}
Krishnateja Killamsetty, Durga Sivasubramanian, Ganesh Ramakrishnan, and
  Rishabh Iyer.
\newblock Glister: Generalization based data subset selection for efficient and
  robust learning.
\newblock In \emph{Proceedings of the AAAI Conference on Artificial
  Intelligence}, 2021{\natexlab{b}}.

\bibitem[Killamsetty et~al.(2021{\natexlab{c}})Killamsetty, Zhao, Chen, and
  Iyer]{killamsetty2021retrieve}
Krishnateja Killamsetty, Xujiang Zhao, Feng Chen, and Rishabh Iyer.
\newblock Retrieve: Coreset selection for efficient and robust semi-supervised
  learning.
\newblock \emph{Advances in Neural Information Processing Systems},
  34:\penalty0 14488--14501, 2021{\natexlab{c}}.

\bibitem[Koh and Liang(2017)]{influence}
Pang~Wei Koh and Percy Liang.
\newblock Understanding black-box predictions via influence functions.
\newblock In \emph{International conference on machine learning}, 2017.

\bibitem[Krizhevsky et~al.(2009)Krizhevsky, Hinton, et~al.]{cifar}
Alex Krizhevsky, Geoffrey Hinton, et~al.
\newblock Learning multiple layers of features from tiny images.
\newblock 2009.

\bibitem[Le and Yang(2015)]{tiny}
Ya~Le and Xuan~S. Yang.
\newblock Tiny imagenet visual recognition challenge.
\newblock 2015.

\bibitem[Li et~al.(2020{\natexlab{a}})Li, Lyu, Koren, Lyu, Li, and Ma]{nad}
Yige Li, Xixiang Lyu, Nodens Koren, Lingjuan Lyu, Bo~Li, and Xingjun Ma.
\newblock Neural attention distillation: Erasing backdoor triggers from deep
  neural networks.
\newblock In \emph{International Conference on Learning Representations},
  2020{\natexlab{a}}.

\bibitem[Li et~al.(2021{\natexlab{a}})Li, Lyu, Koren, Lyu, Li, and Ma]{abl}
Yige Li, Xixiang Lyu, Nodens Koren, Lingjuan Lyu, Bo~Li, and Xingjun Ma.
\newblock Anti-backdoor learning: Training clean models on poisoned data.
\newblock \emph{Advances in Neural Information Processing Systems},
  2021{\natexlab{a}}.

\bibitem[Li et~al.(2020{\natexlab{b}})Li, Wu, Jiang, Li, and
  Xia]{li2020backdoor}
Yiming Li, Baoyuan Wu, Yong Jiang, Zhifeng Li, and Shu-Tao Xia.
\newblock Backdoor learning: A survey.
\newblock \emph{arXiv preprint arXiv:2007.08745}, 2020{\natexlab{b}}.

\bibitem[Li et~al.(2021{\natexlab{b}})Li, Li, Wu, Li, He, and Lyu]{ssba}
Yuezun Li, Yiming Li, Baoyuan Wu, Longkang Li, Ran He, and Siwei Lyu.
\newblock Invisible backdoor attack with sample-specific triggers.
\newblock In \emph{Proceedings of the IEEE/CVF International Conference on
  Computer Vision}, 2021{\natexlab{b}}.

\bibitem[Liu et~al.(2018{\natexlab{a}})Liu, Dolan-Gavitt, and Garg]{FP}
Kang Liu, Brendan Dolan-Gavitt, and Siddharth Garg.
\newblock Fine-pruning: Defending against backdooring attacks on deep neural
  networks.
\newblock In \emph{International Symposium on Research in Attacks, Intrusions,
  and Defenses}, 2018{\natexlab{a}}.

\bibitem[Liu et~al.(2018{\natexlab{b}})Liu, Ma, Aafer, Lee, Zhai, Wang, and
  Zhang]{trojannn}
Yingqi Liu, Shiqing Ma, Yousra Aafer, Wen-Chuan Lee, Juan Zhai, Weihang Wang,
  and Xiangyu Zhang.
\newblock Trojaning attack on neural networks.
\newblock In \emph{25th Annual Network and Distributed System Security
  Symposium}, 2018{\natexlab{b}}.

\bibitem[Mirzasoleiman et~al.(2020)Mirzasoleiman, Bilmes, and
  Leskovec]{mirzasoleiman2020coresets}
Baharan Mirzasoleiman, Jeff Bilmes, and Jure Leskovec.
\newblock Coresets for data-efficient training of machine learning models.
\newblock In \emph{International Conference on Machine Learning}, 2020.

\bibitem[Nguyen and Tran(2020)]{nguyen2020input}
Tuan~Anh Nguyen and Anh Tran.
\newblock Input-aware dynamic backdoor attack.
\newblock \emph{Advances in Neural Information Processing Systems},
  33:\penalty0 3454--3464, 2020.

\bibitem[Nguyen and Tran(2021)]{wanet}
Tuan~Anh Nguyen and Anh~Tuan Tran.
\newblock Wanet - imperceptible warping-based backdoor attack.
\newblock In \emph{International Conference on Learning Representations}, 2021.

\bibitem[Nohyun et~al.(2023)Nohyun, Choi, and Chung]{nohyundata}
Ki~Nohyun, Hoyong Choi, and Hye~Won Chung.
\newblock Data valuation without training of a model.
\newblock In \emph{International Conference on Learning Representations}, 2023.

\bibitem[Paul et~al.(2021)Paul, Ganguli, and Dziugaite]{paul2021deep}
Mansheej Paul, Surya Ganguli, and Gintare~Karolina Dziugaite.
\newblock Deep learning on a data diet: Finding important examples early in
  training.
\newblock \emph{Advances in Neural Information Processing Systems},
  34:\penalty0 20596--20607, 2021.

\bibitem[Rumelhart et~al.(1986)Rumelhart, Hinton, and Williams]{BP}
David~E Rumelhart, Geoffrey~E Hinton, and Ronald~J Williams.
\newblock Learning representations by back-propagating errors.
\newblock \emph{Nature}, 323\penalty0 (6088):\penalty0 533--536, 1986.

\bibitem[Salem et~al.(2022)Salem, Wen, Backes, Ma, and Zhang]{salem2022dynamic}
Ahmed Salem, Rui Wen, Michael Backes, Shiqing Ma, and Yang Zhang.
\newblock Dynamic backdoor attacks against machine learning models.
\newblock In \emph{IEEE 7th European Symposium on Security and Privacy
  (EuroS\&P)}, 2022.

\bibitem[Sener and Savarese(2017)]{sener2017active}
Ozan Sener and Silvio Savarese.
\newblock Active learning for convolutional neural networks: A core-set
  approach.
\newblock \emph{arXiv preprint arXiv:1708.00489}, 2017.

\bibitem[Souri et~al.(2021)Souri, Goldblum, Fowl, Chellappa, and
  Goldstein]{souri2021sleeper}
Hossein Souri, Micah Goldblum, Liam Fowl, Rama Chellappa, and Tom Goldstein.
\newblock Sleeper agent: Scalable hidden trigger backdoors for neural networks
  trained from scratch.
\newblock \emph{arXiv preprint arXiv:2106.08970}, 2021.

\bibitem[Toneva et~al.(2018)Toneva, Sordoni, Combes, Trischler, Bengio, and
  Gordon]{forgetting}
Mariya Toneva, Alessandro Sordoni, Remi Tachet~des Combes, Adam Trischler,
  Yoshua Bengio, and Geoffrey~J Gordon.
\newblock An empirical study of example forgetting during deep neural network
  learning.
\newblock \emph{arXiv preprint arXiv:1812.05159}, 2018.

\bibitem[Tran et~al.(2018)Tran, Li, and Madry]{tran2018spectral}
Brandon Tran, Jerry Li, and Aleksander Madry.
\newblock Spectral signatures in backdoor attacks.
\newblock \emph{Advances in Neural Information Processing Systems}, 31, 2018.

\bibitem[Turner et~al.(2019)Turner, Tsipras, and Madry]{turner2019label}
Alexander Turner, Dimitris Tsipras, and Aleksander Madry.
\newblock Label-consistent backdoor attacks.
\newblock \emph{arXiv preprint arXiv:1912.02771}, 2019.

\bibitem[Wang et~al.(2023{\natexlab{a}})Wang, Chen, Zhu, Liu, Zhang, Fan, and
  Wu]{wang2023robust}
Ruotong Wang, Hongrui Chen, Zihao Zhu, Li~Liu, Yong Zhang, Yanbo Fan, and
  Baoyuan Wu.
\newblock Robust backdoor attack with visible, semantic, sample-specific, and
  compatible triggers.
\newblock \emph{arXiv preprint arXiv:2306.00816}, 2023{\natexlab{a}}.

\bibitem[Wang et~al.(2022)Wang, Zhai, and Ma]{wang2022bppattack}
Zhenting Wang, Juan Zhai, and Shiqing Ma.
\newblock Bppattack: Stealthy and efficient trojan attacks against deep neural
  networks via image quantization and contrastive adversarial learning.
\newblock In \emph{Proceedings of the IEEE/CVF Conference on Computer Vision
  and Pattern Recognition}, 2022.

\bibitem[Wang et~al.(2023{\natexlab{b}})Wang, Mei, Zhai, and
  Ma]{wang2023unicorn}
Zhenting Wang, Kai Mei, Juan Zhai, and Shiqing Ma.
\newblock {UNICORN}: A unified backdoor trigger inversion framework.
\newblock In \emph{International Conference on Learning Representations},
  2023{\natexlab{b}}.

\bibitem[Wang et~al.(2004)Wang, Bovik, Sheikh, and Simoncelli]{wang2004image}
Zhou Wang, Alan~C Bovik, Hamid~R Sheikh, and Eero~P Simoncelli.
\newblock Image quality assessment: from error visibility to structural
  similarity.
\newblock \emph{IEEE transactions on image processing}, 13\penalty0
  (4):\penalty0 600--612, 2004.

\bibitem[Wu et~al.(2022)Wu, Chen, Zhang, Zhu, Wei, Yuan, and
  Shen]{wu2022backdoorbench}
Baoyuan Wu, Hongrui Chen, Mingda Zhang, Zihao Zhu, Shaokui Wei, Danni Yuan, and
  Chao Shen.
\newblock Backdoorbench: A comprehensive benchmark of backdoor learning.
\newblock In \emph{Thirty-sixth Conference on Neural Information Processing
  Systems Datasets and Benchmarks Track}, 2022.

\bibitem[Wu et~al.(2023)Wu, Liu, Zhu, Liu, He, and Lyu]{wu2023adversarial}
Baoyuan Wu, Li~Liu, Zihao Zhu, Qingshan Liu, Zhaofeng He, and Siwei Lyu.
\newblock Adversarial machine learning: A systematic survey of backdoor attack,
  weight attack and adversarial example.
\newblock \emph{arXiv preprint arXiv:2302.09457}, 2023.

\bibitem[Xia et~al.(2022)Xia, Li, Zhang, and Li]{ijcai2022forget}
Pengfei Xia, Ziqiang Li, Wei Zhang, and Bin Li.
\newblock Data-efficient backdoor attacks.
\newblock In \emph{Proceedings of the Thirty-First International Joint
  Conference on Artificial Intelligence}, 2022.

\bibitem[Yoon et~al.(2020)Yoon, Arik, and Pfister]{yoon2020data}
Jinsung Yoon, Sercan Arik, and Tomas Pfister.
\newblock Data valuation using reinforcement learning.
\newblock In \emph{International Conference on Machine Learning}, 2020.

\bibitem[Zeng et~al.(2021{\natexlab{a}})Zeng, Chen, Park, Mao, Jin, and
  Jia]{ibau}
Yi~Zeng, Si~Chen, Won Park, Zhuoqing Mao, Ming Jin, and Ruoxi Jia.
\newblock Adversarial unlearning of backdoors via implicit hypergradient.
\newblock In \emph{International Conference on Learning Representations},
  2021{\natexlab{a}}.

\bibitem[Zeng et~al.(2021{\natexlab{b}})Zeng, Park, Mao, and
  Jia]{zeng2021rethinking}
Yi~Zeng, Won Park, Z~Morley Mao, and Ruoxi Jia.
\newblock Rethinking the backdoor attacks' triggers: A frequency perspective.
\newblock In \emph{Proceedings of the IEEE/CVF International Conference on
  Computer Vision}, pages 16473--16481, 2021{\natexlab{b}}.

\bibitem[Zeng et~al.(2022)Zeng, Pan, Just, Lyu, Qiu, and
  Jia]{zeng2022narcissus}
Yi~Zeng, Minzhou Pan, Hoang~Anh Just, Lingjuan Lyu, Meikang Qiu, and Ruoxi Jia.
\newblock Narcissus: A practical clean-label backdoor attack with limited
  information.
\newblock \emph{arXiv preprint arXiv:2204.05255}, 2022.

\bibitem[Zeng et~al.(2023)Zeng, Pan, Jahagirdar, Jin, Lyu, and
  Jia]{zeng2023sift}
Yi~Zeng, Minzhou Pan, Himanshu Jahagirdar, Ming Jin, Lingjuan Lyu, and Ruoxi
  Jia.
\newblock How to sift out a clean data subset in the presence of data
  poisoning?
\newblock In \emph{USENIX Security Symposium}, 2023.

\bibitem[Zhang et~al.(2022)Zhang, Dongdong, Huang, Liao, Zhang, Feng, Hua, and
  Yu]{zhang2022poison}
Jie Zhang, Chen Dongdong, Qidong Huang, Jing Liao, Weiming Zhang, Huamin Feng,
  Gang Hua, and Nenghai Yu.
\newblock Poison ink: Robust and invisible backdoor attack.
\newblock \emph{IEEE Transactions on Image Processing}, 31:\penalty0
  5691--5705, 2022.

\bibitem[Zheng et~al.(2022)Zheng, Tang, Li, and Liu]{clp}
Runkai Zheng, Rongjun Tang, Jianze Li, and Li~Liu.
\newblock Data-free backdoor removal based on channel lipschitzness.
\newblock In \emph{European Conference on Computer Vision}, 2022.

\bibitem[Zhu et~al.(2023)Zhu, Wei, Shen, Fan, and Wu]{zhu2023enhancing}
Mingli Zhu, Shaokui Wei, Li~Shen, Yanbo Fan, and Baoyuan Wu.
\newblock Enhancing fine-tuning based backdoor defense with sharpness-aware
  minimization.
\newblock \emph{arXiv preprint arXiv:2304.11823}, 2023.

\end{thebibliography}
